\documentclass[fleqn,usenatbib]{mnras}

\usepackage{graphicx}	
\usepackage{amsmath}	
\usepackage{multicol}        
\usepackage{bm}		
\usepackage{pdflscape}	

\usepackage{latexsym}
\usepackage{longtable,booktabs}
\usepackage{caption,rotating}
\usepackage{scalerel}

\usepackage[T1]{fontenc}
\usepackage{ae,aecompl}
\usepackage{newtxtext,newtxmath}

\DeclareRobustCommand{\VAN}[3]{#2}
\let\VANthebibliography\thebibliography
\def\thebibliography{\DeclareRobustCommand{\VAN}[3]{##3}\VANthebibliography}

\title[Identifying Type IL GRBs based on ML]{Identifying Merger-Driven Long Gamma-Ray Bursts based on Machine Learning}

\author[Zhu et al.]{Si-Yuan Zhu$^{1,2}$, Hui-Ying Deng$^{1,2}$, Fu-Wen Zhang$^{3,4}$, Qian-Zi Mo$^{1,2}$, Pak-Hin Thomas Tam$^{1,2}$%
	\thanks{Contact e-mail: \href{mailto:tanbxuan@mail.sysu.edu.cn}{tanbxuan@mail.sysu.edu.cn}}%
	\\
	$^{1}$School of Physics and Astronomy, Sun Yat-Sen University, Zhuhai 519082, China\\
	$^{2}$CSST Science Center for the Guangdong-Hong Kong-Macau Greater Bay Area, Sun Yat-sen University, Zhuhai 519082, China\\
	$^{3}$College of Physics and Electronic Information Engineering, Guilin University of Technology, Guilin 541004, China\\
	$^{4}$Key Laboratory of Low-dimensional Structural Physics and Application, Education Department of Guangxi Zhuang Autonomous\\ Region, Guilin 541004, China}

\date{Last updated xx xx xx; in original form xx xx xx}

\pubyear{2025}

\begin{document}
\label{firstpage}
\pagerange{\pageref{firstpage}--\pageref{lastpage}}
\maketitle

\begin{abstract}
Gamma-ray bursts (GRBs) are classified as Type I GRBs originated from compact binary mergers and Type II GRBs originated from massive collapsars.
While Type I GRBs are typically shorter than 2 seconds, recent observations suggest that some extend to tens of seconds, forming a potential subclass, Type IL GRBs.
However, apart from their association with kilonovae, so far no rapid identification is possible.
Given the uncertainties and limitations of optical and infrared afterglow observations, an identification method based solely on prompt emission can make such identification possible for many more GRBs. 
Interestingly, two established Type IL GRBs: GRB 211211A and GRB 230307A, exhibit a three-episode structure: precursor emission (PE), main emission (ME), and extended emission.
Therefore, we comprehensively search for GRBs in the Fermi/GBM catalog and identify 29 three-episode GRBs.
Based on 12 parameters, we utilize machine learning to distinguish Type IL GRBs from Type II GRBs.
Apart from GRB 211211A and GRB 230307A, we are able to identify six more previously unknown Type IL GRBs: GRB 090831, GRB 170228A, GRB 180605A, GRB 200311A, GRB 200914A, and GRB 211019A. 
We find that Type IL GRBs are characterized by short duration and minimum variability timescale of PE, a short waiting time between PE and ME, and that ME follows the $E_{\rm p,z}$--$E_{\rm iso}$ correlation of Type I GRBs.
For the first time, we identify a high-significant PE in the confirmed Type IL GRB 060614.
\end{abstract}


\section{Introduction} \label{sec:introduction}
Gamma-ray bursts (GRBs) are extremely energetic explosions occurring on the cosmological scale.
Previously, observations have revealed that GRBs are associated with Type Ic supernovae (SNe) or gravitational waves (GWs)/kilonovae (KNe).
The association between GRBs and SNe suggests that a part of GRB progenitors are massive collapsars \citep[Type II GRBs;][]{1993ApJ...405..273W,1998Natur.395..670G,2006ARA&A..44..507W}.
The association between GRB and GW/KN suggests that a part of GRB progenitors are compact binary mergers \citep[Type I GRBs;][]{1986ApJ...308L..43P,1992ApJ...395L..83N,2017PhRvL.119p1101A,2017ApJ...851L..18W}.

Based on the bimodal distribution of prompt emission durations, GRBs are traditionally classified into long GRBs (LGRBs; $T_{90} > 2$ s) and short GRBs \citep[SGRBs; $T_{90} < 2$ s;][] {1993ApJ...413L.101K}.
Previously, theories and observations support that LGRBs and SGRBs originate from the collapsars and mergers, respectively \citep{1998Natur.395..670G,1999ApJ...524..262M,2005A&A...436..273A,2005Natur.437..851G}.
Phenomenologically, SGRBs and LGRBs exhibit different properties.
Compared to LGRBs, SGRBs generally have negligible or negative spectral lag \citep[$\tau$;][]{2006MNRAS.367.1751Y,2015MNRAS.446.1129B}, shorter minimum variability timescale \citep[MVT;][]{2013MNRAS.432..857M,2023A&A...671A.112C}, larger spectral hardness \citep{2012ApJ...750...88Z}, and a distinct correlation between the isotropic energy and the peak energy in the rest frame \citep[$E_{\rm p,z}$--$E_{\rm iso}$ correlation;][]{2002A&A...390...81A,2013MNRAS.430..163Q,2023ApJ...950...30Z}.

Recent observations have revealed that several LGRBs, including GRB 211211A and GRB 230307A, are associated with KNe \citep[i.e., Type IL GRBs;][]{2022Natur.612..223R,2022Natur.612..232Y,2024Natur.626..742Y,2024Natur.626..737L}, while a SGRB, GRB 200826A, is associated with a SN \citep{2021NatAs...5..917A,2021NatAs...5..911Z}, which suggests that the dichotomy based solely on $T_{90}$ duration does not effectively differentiate progenitors from distinct populations.
Interestingly, the whole emission (WE) of these Type IL GRB light curves can be distinctly divided into three episodes: precursor emission (PE), main emission (ME), and extended emission (EE) episodes, which are separated by distinct quiescent episodes \citep[i.e., waiting time;][]{2024ApJ...969...26P,2025ApJ...979...73W}. 
These separated episodes are not only a unique observational feature, but may also correspond to different emission mechanisms or progenitor activities, providing a new perspective on the complex physical processes within GRBs.
Interestingly, despite their main emission longer than 2 s, they share similar observational properties with SGRBs, while their whole emissions are more similar to that of LGRBs \citep{2022ApJ...936L..10Z,2024ApJ...969...26P}.
This dual nature suggests that Type IL GRBs may represent a transitional subclass between traditional SGRBs and LGRBs, further blurring the distinction between these populations.

These observations raise an important question about whether the three-episode structure reveal intrinsic properties of Type IL GRBs. 
Additionally, it is worth investigating whether there are subtle differences between the prompt emission of GRBs from different progenitors, especially when the duration of GRBs alone cannot reliably indicate their origins.
If such differences exist, which emission episode could provide clues to distinguish them?
Unfortunately, both the precursor emission and extended emission are relatively weak, with observed occurrence rates below 30\%, and even fewer are detected simultaneously \citep{2015MNRAS.452..824K,2020PhRvD.102j3014C}.
As a result, our understanding of their origins, physical mechanisms, and interconnections remains highly limited, with significant controversies persisting in these aspects.
A comprehensive study of three-episode GRBs could greatly enhance our understanding of the physical processes involved. This may provide key clues to uncovering the progenitor, jet composition, and radiation mechanism of GRBs.

In this work, we present the first comprehensively search for three-episode GRBs in the Fermi/GBM catalog and compile a three-episode GRB sample. 
We note that both mergers and collapsars can produce such long-duration three-episode GRBs. 
Distinguishing these two progenitors based on prompt emission has long been a topic of interest.
Therefore, we aim to apply machine learning techniques, which have been widely used in GRB classification, in attempts to distinguish between them \citep[see][and the references therein]{2024MNRAS.532.1434Z}.
The t-distributed stochastic neighbor embedding \citep[t-SNE;][]{2008JMLR.9.2579M,2014JMLR.15.3221M} and the Uniform Manifold Approximation and Projection \citep[UMAP;][]{2018arXiv180203426M} are two powerful unsupervised dimensionality reduction algorithms.
These methods effectively map neighboring data points from high-dimensional space to two-dimensional space, facilitating GRB classification based on light curves or physical parameters without the need for prior information.
Additionally, we individually analyze and compare the temporal and spectral properties of their precursor, main, extended, and whole emissions, and discuss their possible origins.

The structure of this paper is organized as follows.
In Section \ref{sec:data}, we describe the sample selection criteria and data analysis methods.
In Section \ref{sec:classification}, we use machine learning to distinguish between typical Type II GRBs and peculiar long-duration Type I GRBs.
Discussion are presented in Section \ref{sec:discussions}, and conclusions are given in Section \ref{sec:conclusions}.
The symbolic notation $Q_{\rm n} = Q/10^{\rm n}$ is adopted.

\section{Data Selection and Analysis} \label{sec:data}

The Fermi satellite, launched in 2008, has collected an extensive dataset of GRB observations with spectral coverage ranging from 8 keV--300 GeV.
In this work, we only used the time-tagged event (TTE) data of Gamma-ray Burst Monitor (GBM) for temporal and spectral analysis with an effective energy range of 8 keV--40 MeV \citep{2009ApJ...702..791M}.

\subsection{Light Curve Extraction and Temporal Analysis}\label{subsec:lightcurve}
We comprehensively searched for GRBs in the Fermi/GBM catalog from August 2008 to December 2024 \citep{2020ApJ...893...46V}.
We then extracted the light curves of GRBs in the 8--1000 keV band using the standard fermi tool for Python source \texttt{GBM Data Tools}.
The Bayesian block method is a powerful tool for the non-parametric analysis of GRB light curves, effectively identifying and characterizing statistically significant changes \citep{2013ApJ...764..167S}.
In this work, we apply the Bayesian block method to the light curves to determine the optimal change points, segmenting the whole emission into distinct episodes of precursor emission, main emission, and extended emission, as well as the quiescent episodes between the precursor and main emissions and between the main and extended emissions.
For the Bayesian block method, the false-positive rate for a given change point is set to 0.05.

\begin{figure*}
	\centering
	\includegraphics[angle=0,scale=0.34]{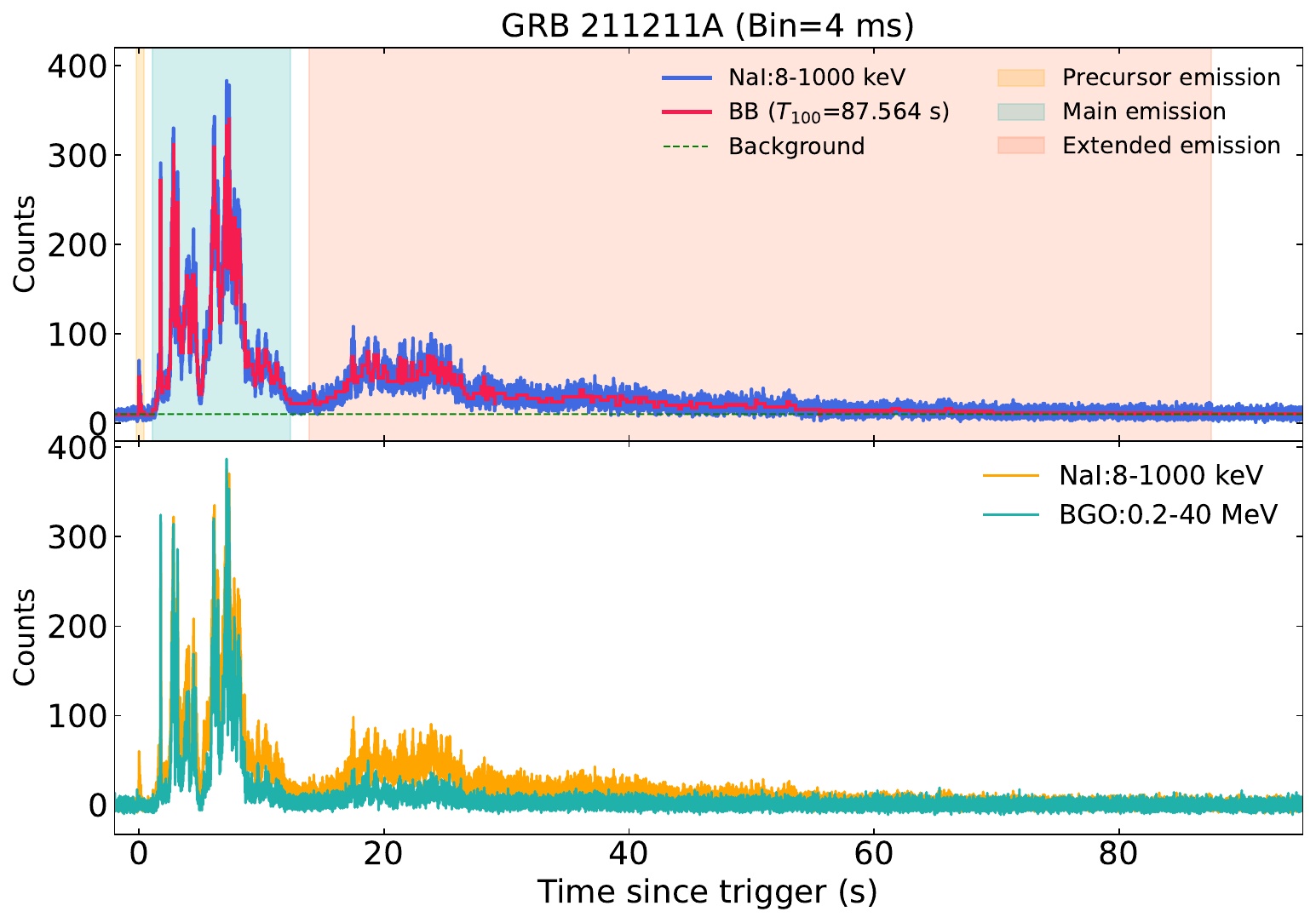}
	\includegraphics[angle=0,scale=0.34]{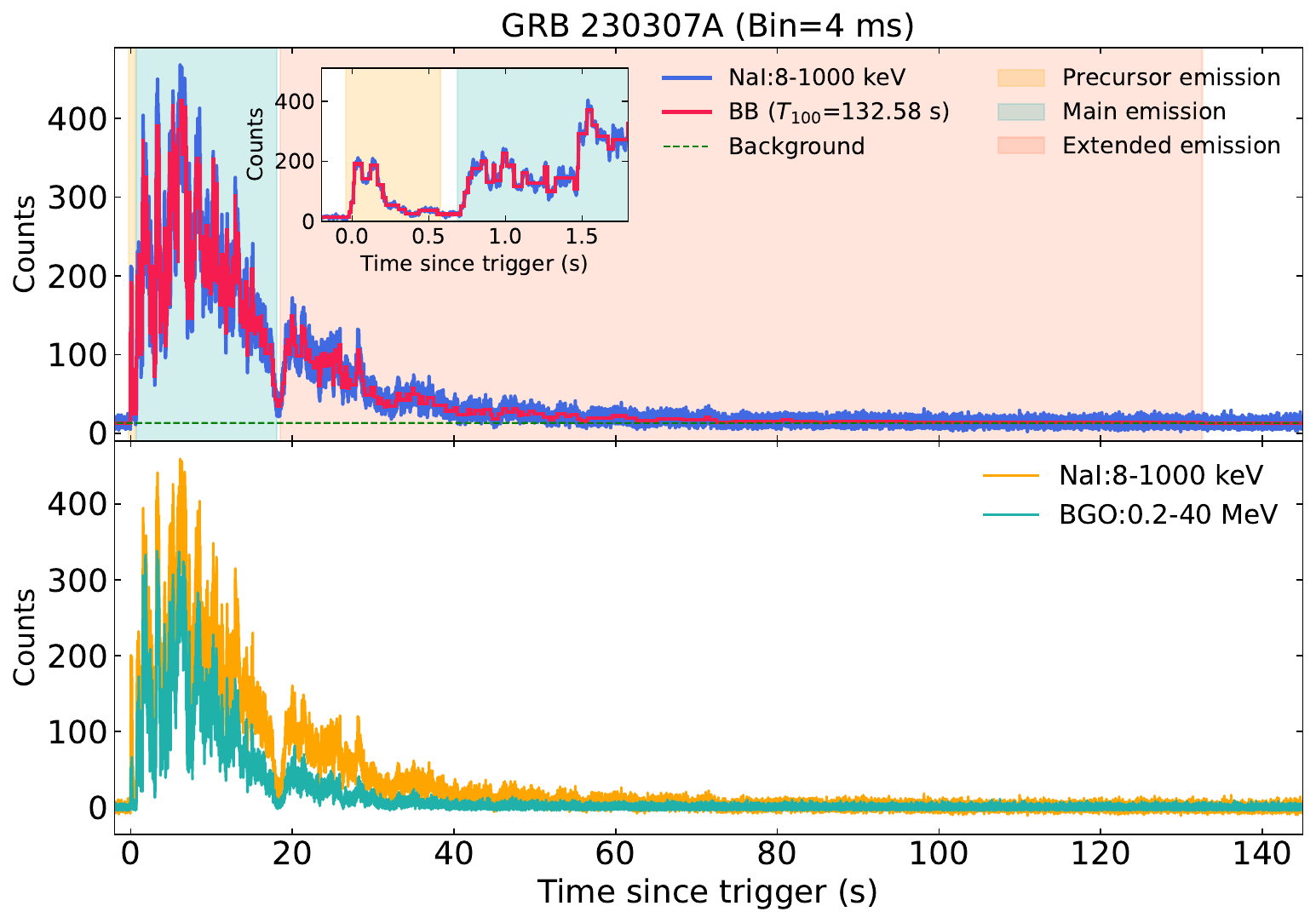}
	\includegraphics[angle=0,scale=0.34]{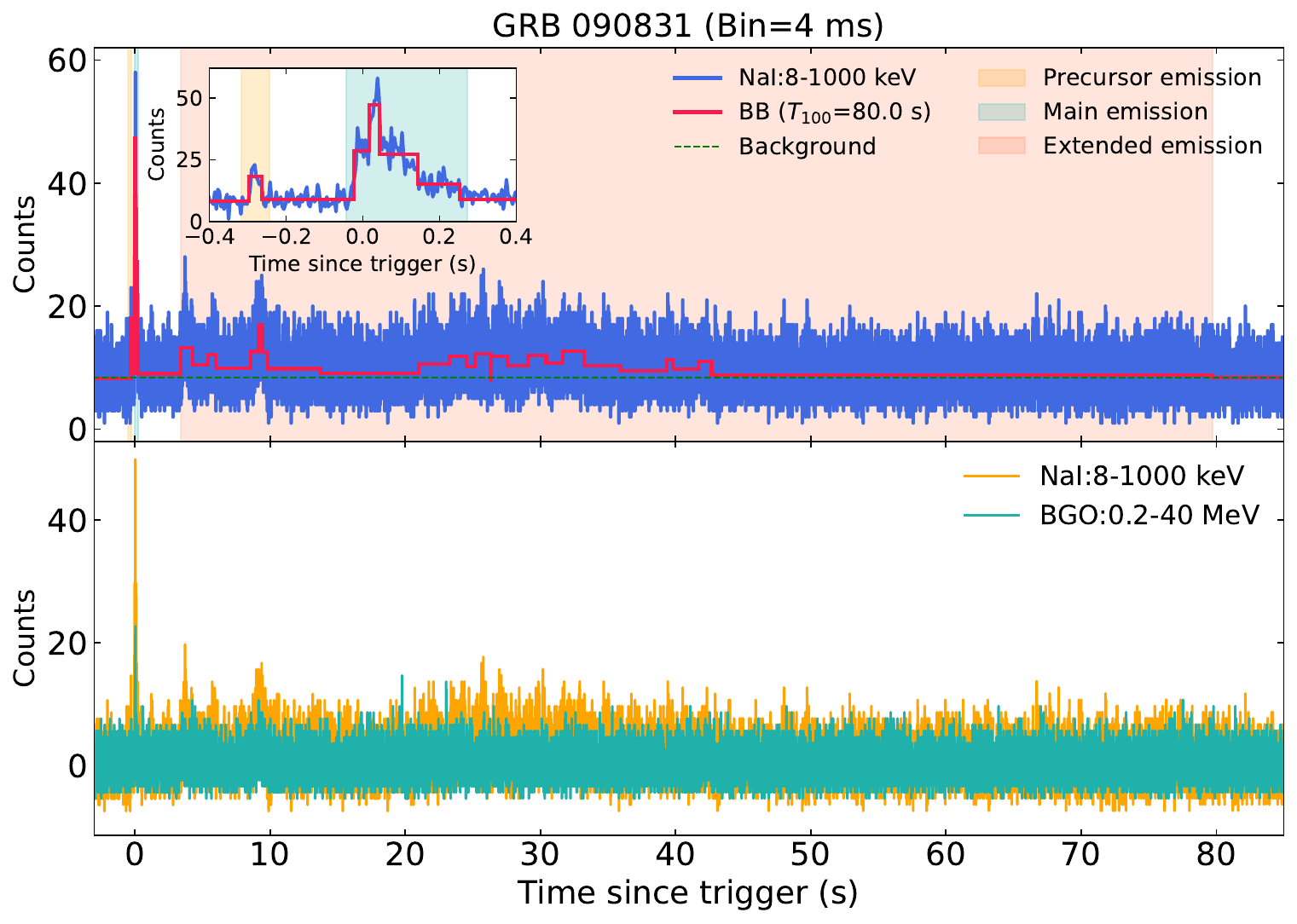}
	\includegraphics[angle=0,scale=0.34]{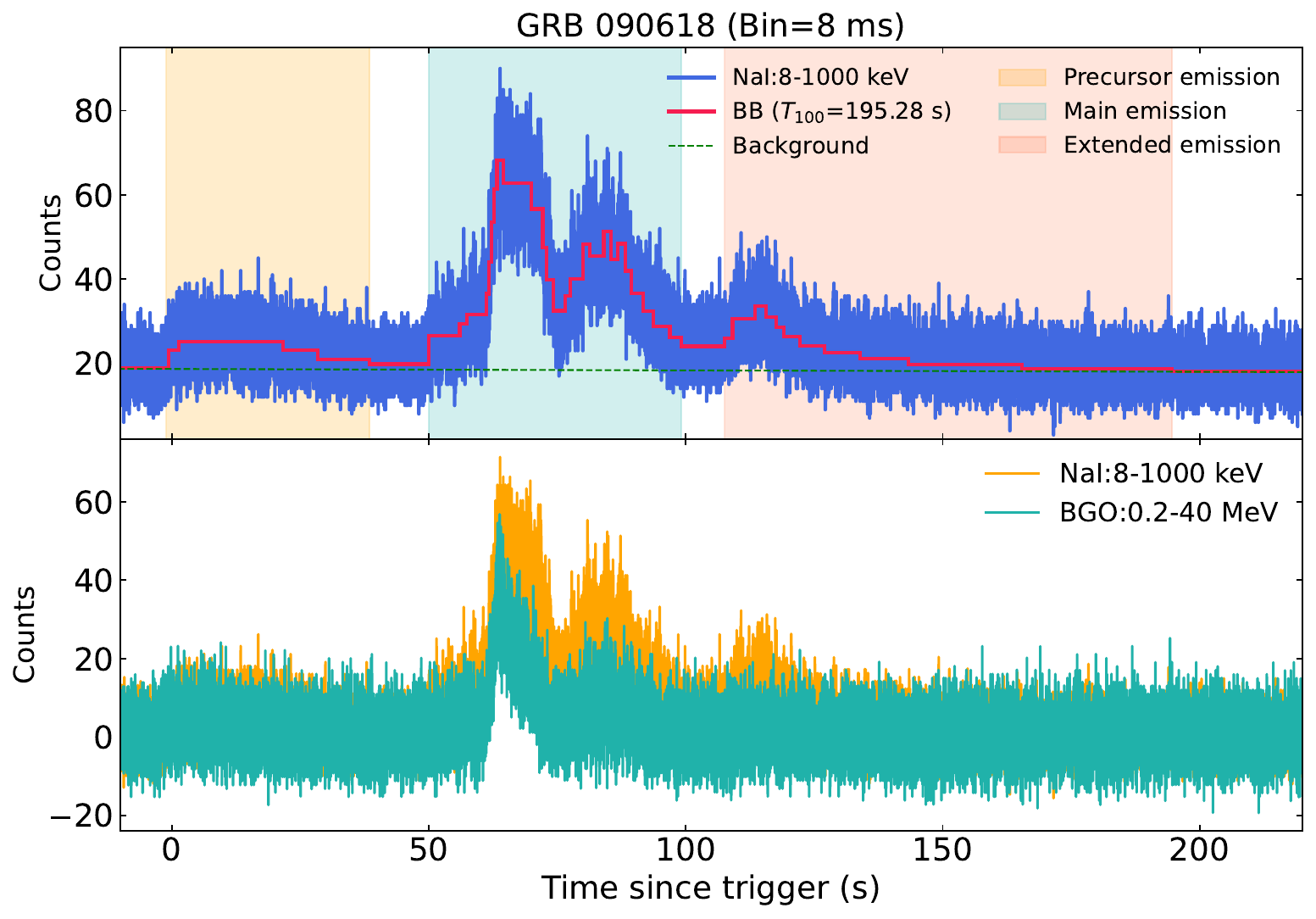}
	\caption{The light curves of the three-episode GRBs. The blue lines are the light curve in the 8--1000 keV range. The red lines are the Bayesian blocks. The orange and green lines are the light curves observed by NaI and BGO detector, respectively. The green dashed line is the background. The yellow, lightcyan, and red shaded intervals represent precursor, main, and extended emissions, respectively. The zoomed-in result for the overlapping part of the precursor and main emissions in the figure is displayed in the inset.}
	\label{figure:lightcurve}
\end{figure*}

The identification of precursor emission and extended emission is based exclusively on light curves.
The criteria for precursor emission are that it occurs before the main emission, exhibits prominent pulse features in the Bayesian block, has an intensity weaker than that of the main emission, and is separated from the main emission by a distinct quiescent episode.
The criteria for extended emission are that it must occur after the main emission and be preceded by a noticeable quiescent episode.
The analysis of the precursor emission in GRB 230307A suggests that the weak emission during the quiescent episode may be affected by redshift effects \citep{2025ApJ...979...73W}. 
Therefore, the flux during the quiescent episode is not required to drop to the background level but can be characterized by stable emission close to the background.
In other words, the quiescent episode is defined as a steady emission episode occurring after the decay of the preceding episode and before the rapid rise of the subsequent episode.
The duration of quiescent episode should be at least 10 times the time resolution to minimize random coincidences caused by statistical fluctuations. 
For the quiescent episode between the precursor emission and the main emission, a signal-to-noise ratio (SNR) of less than 3 $\sigma$ is required (3.15 $\sigma$ for GRB 230307A).
In contrast, no specific SNR threshold is imposed for the quiescent episode between the main emission and the extended emission.
In total, we identified 29 three-episode GRBs that meet these criteria from 3883 GRBs.
Examples of such GRBs are shown in Figure \ref{figure:lightcurve}.
Note that, because initial screening relies on the eye, there may be omissions and some low SNR GRBs were also excluded.

We analyze the light curves of the precursor, main, extended, and quiescent emission episodes, respectively.
We calculated the durations of the precursor emission ($T_{\rm 100,PE}$), main emission ($T_{\rm 100,ME}$), extended emission ($T_{\rm 100,EE}$), whole emission ($T_{\rm 100,WE}$), as well as the quiescent episodes between the precursor and main emissions ($T_{\rm wt1}$), and between the main and extended emissions ($T_{\rm wt2}$).
The duration of an emission episode is defined as the difference between the end time and the start time of the episode.

The spectral lag ($\tau$) is defined as the time delay of low-energy photons with respect to high-energy photons \citep{1986ApJ...301..213N,1995A&A...300..746C}.
Generally, $\tau$ of LGRBs or Type II GRBs are significant, whereas those of SGRBs or Type I GRBs are negligible or negative.
Although Type II GRBs can also exhibit negative $\tau$, $\tau$ remains a key indicator of GRB origins.

We used the cross-correlation function (CCF) to calculate the spectral lag of the 50--100 keV relative to the 25--50 keV ($\tau_{32}$) and the 100--200 keV relative to the 25--50 keV ($\tau_{42}$).
Due to the potential impact of low SNR on $\tau$ calculations, $\tau$ is not computed if the CCF cannot be reliably fitted at a time resolution below 256 ms.

The MVT represents the shortest time scale at which changes in a GRB light curve can be identified.
Generally, the MVT of LGRBs is longer than that of SGRBs. 
However, relatively short MVTs have also been identified in Type IL GRBs, which may serve as a distinguishing feature of them from typical LGRBs \citep{2023ApJ...954L...5V,2025ApJ...979...73W}. It has been found that MVTs typically corresponds to the rise time of the shortest pulse in the light curve \citep{2014ApJ...787...90G,2015ApJ...811...93G,2025ApJ...979...73W}.
In this work, we use Bayesian block method to identify the shortest pulse in the precursor, main, and extended emissions, and refer it as $T_{\rm MVT,PE}$, $T_{\rm MVT,ME}$, and $T_{\rm MVT,EE}$, respectively \citep{2023ApJS..268....5X,2025ApJ...979...73W}.
The results of the temporal analysis are listed in Table \ref{table:temporal}.

\begin{table*}
	\caption{The temporal properties of three-episode GRBs}
	\label{table:temporal}
	\scriptsize
	\setlength{\tabcolsep}{8pt}
	\renewcommand{\arraystretch}{1.3}
	\begin{tabular}{lccccccccccc}
		\hline
		GRB & $T_{\rm res}$ &$T_{\rm wt1}$ & $T_{\rm wt2}$ & Episode & $T_{\rm 100,start}$ & $T_{\rm 100,end}$ & $T_{100}$ & $T_{\rm MVT}$ & $\tau_{\rm res}$ & $\tau_{32}$ & $\tau_{42}$ \\
		& (ms) & (s) & (s) &  & (s) & (s) & (s) & (s) & (ms) & (ms) & (ms)\\
		\hline
		090131 & 4 & 1.78 & 12.696 & PE & -0.276 & 0.316 & 0.592 & 0.102 & 128 & $-170.07 \pm 255.51$ & $16.22 \pm 77.51$ \\
		& & & & ME & 2.096 & 9.78 & 7.684 & 0.974 & 8 & $80.99 \pm 4.71$ & $143.41 \pm 24.26$ \\
		& & & & EE & 22.476 & 84.04 & 61.564 & 0.594 & 8 & $78.13 \pm 4.42$ & $119.36 \pm 19.84$ \\
		& & & & WE & -0.276 & 84.04 & 84.316 & 0.102 & 8 & $70.66 \pm 2.22$ & $120.58 \pm 41.24$ \\
		\hline
		090618 & 8 & 11.488 & 8.472 & PE & -0.632 & 38.504 & 39.136 & 1.644 & 64 & $3282.52 \pm 21.25$ & $5474.62 \pm 0.62$ \\
		& & & & ME & 49.992 & 99.144 & 49.152 & 4.284 & 64 & $428.09 \pm 9.10$ & $791.97 \pm 4.64$ \\
		& & & & EE & 107.616 & 194.648 & 87.032 & 7.244 & 64 & $741.04 \pm 56.4$ & $2380.64 \pm 44.13$ \\
		& & & & WE & -0.632 & 194.648 & 195.28 & 1.644 & 64 & $405.16 \pm 2.75$ & $797.51 \pm 7.28$ \\
		\hline
		\multicolumn{12}{l}{This table is published in its entirety in the machine-readable format. A full machine-readable version is available online.}\\
	\end{tabular}
\end{table*}

\subsection{Spectral Analysis and Properties}\label{subsec:rest}
We fit the spectra of the precursor emission, main emission, extended emission, and whole emission for each GRB.
For individual episodes for each GRB, the spectral fitting is conducted using the \texttt{PyXspec} based on the standard \texttt{HEASOFT}.
The Markov Chain Monte Carlo (MCMC) method was used for spectral fitting, with PGSTAT was adopted as the fit statistic.
We use the built-in function of \texttt{PyXspec} to implement the MCMC method. 
The uncertainties of the fitting are typically provided at the 68\% confidence level and are calculated using the last 50\% of $10^{5}$ iterations.
The spectral energy range of 10--900 keV is selected for the NaI detectors and 300 keV--40 MeV for the BGO detectors.
In order to avoid the iodine K-edge at 33.17 keV caused by the NaI detectors, we exclude the energy range of 30--40 keV for the NaI detectors \citep{2009ApJ...702..791M}.
Additionally, since the spectrum of the precursor emission is generally softer, we select a spectral energy range of 300 keV--10 MeV for the BGO detectors when analyzing precursor emissions to minimize the influence of high-energy data on the selection of the best-fit model.

The selection of the best model is based on the Bayesian Information Criterion (BIC), defined as $\text{BIC} = -2 \ln L + k \ln N$, where $L$ is the maximum likelihood, $k$ is the number of free parameters in the model, and $N$ is the number of data points.
The model with a lower BIC value is preferred, especially when the difference in BIC ($\Delta \rm BIC$) larger than 10.
Additionally, the best-fit model should have all parameters well constrained.

The basic spectrum model used in our analysis including Band \citep{1997ApJ...486..928B}, cutoff power-law (CPL), power-law (PL), and blackbody (BB) models, which are the most commonly used models for GRB spectral fitting.
The PL model is defined as
\begin{equation}\label{PL}
	N(E) = A\left( \frac{E}{1\,\mathrm{keV}} \right)^\alpha,
\end{equation}
where $\alpha$ is the photon index, and $A$ is the normalization constant.
The CPL function is defined as
\begin{equation}\label{CPL}
	N(E) = A\left( \frac{E}{1\,\mathrm{keV}} \right)^\alpha \exp\left( -\frac{E}{E_0} \right)
\end{equation}
where $\alpha$ is the photon index, $E_0$ is the characteristic cutoff energy in units of keV, and $A$ is the normalization constant.
The Band model is defined as
\begin{equation}\label{Band}
	N(E) = 
	\begin{cases} 
		A\left( \frac{E}{100\,\mathrm{keV}} \right)^\alpha 
		\exp\left( -\frac{E}{E_0} \right), & E < (\alpha - \beta)E_0, \\[8pt]
		\begin{aligned}
			& \textstyle A\left( \frac{E}{100\,\mathrm{keV}} \right)^\beta 
			\left[ \frac{(\alpha - \beta)E_0}{100\,\mathrm{keV}} \right]^{\alpha-\beta} \\[4pt]
			& \textstyle \qquad \times \exp(\beta - \alpha),
		\end{aligned} & E \geq (\alpha - \beta)E_0,
	\end{cases}
\end{equation}
where $\alpha$ is the low energy spectral index, $\beta$ is the high energy spectral index, $E_{0}$ is the characteristic cutoff energy in units of keV, and $A$ is the normalization constant.
For the CPL and Band model, the peak energy in the $\nu f_\nu$ spectrum is expressed as $E_{\rm p} = E_{0}(2+\alpha)$.
The BB model is defined as
\begin{equation}\label{BB}
	N(E) = \frac{1.0344 \times 10^{-3} \times AE^2}{\exp{E/kT} - 1}
\end{equation}
where $kT$ is the temperature in units of keV, $k$ is the Boltzmann constant, and $A$ is the normalization constant.
Among them, two GRBs have too faint precursor emissions, and five GRBs have too faint extended emissions for their spectra to be fit. 

\begin{figure*}
	\centering
	\includegraphics[angle=0,scale=0.40]{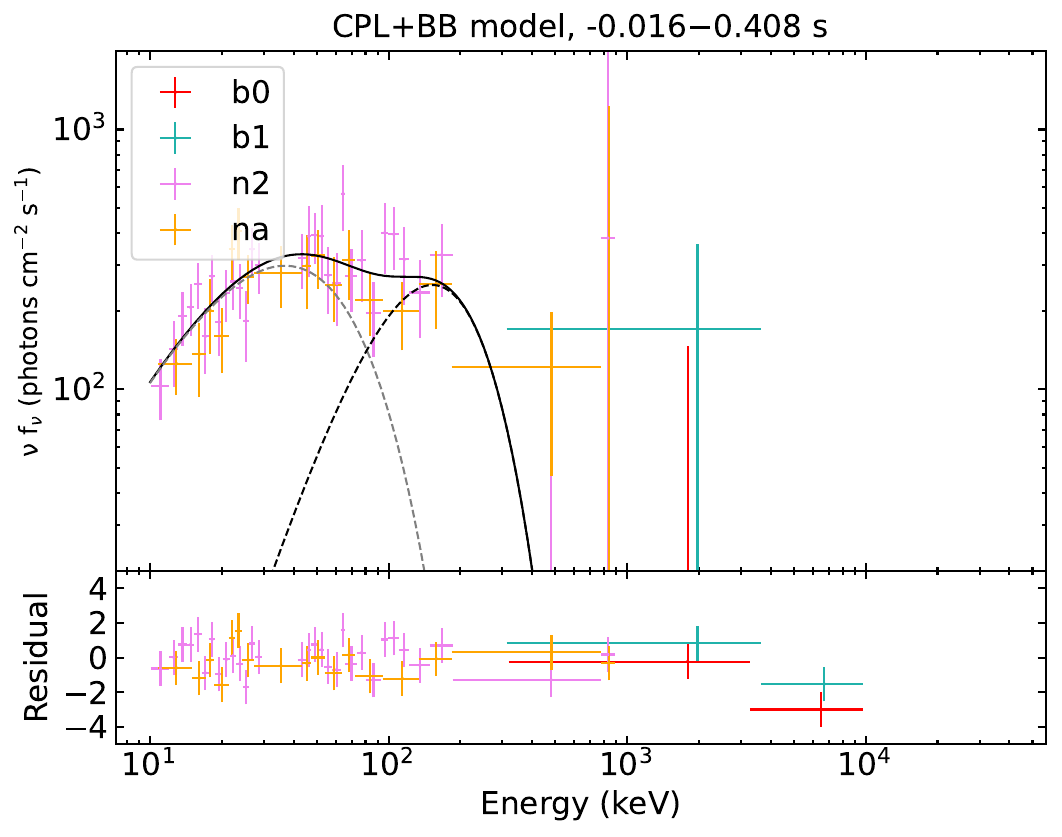}
	\includegraphics[angle=0,scale=0.19]{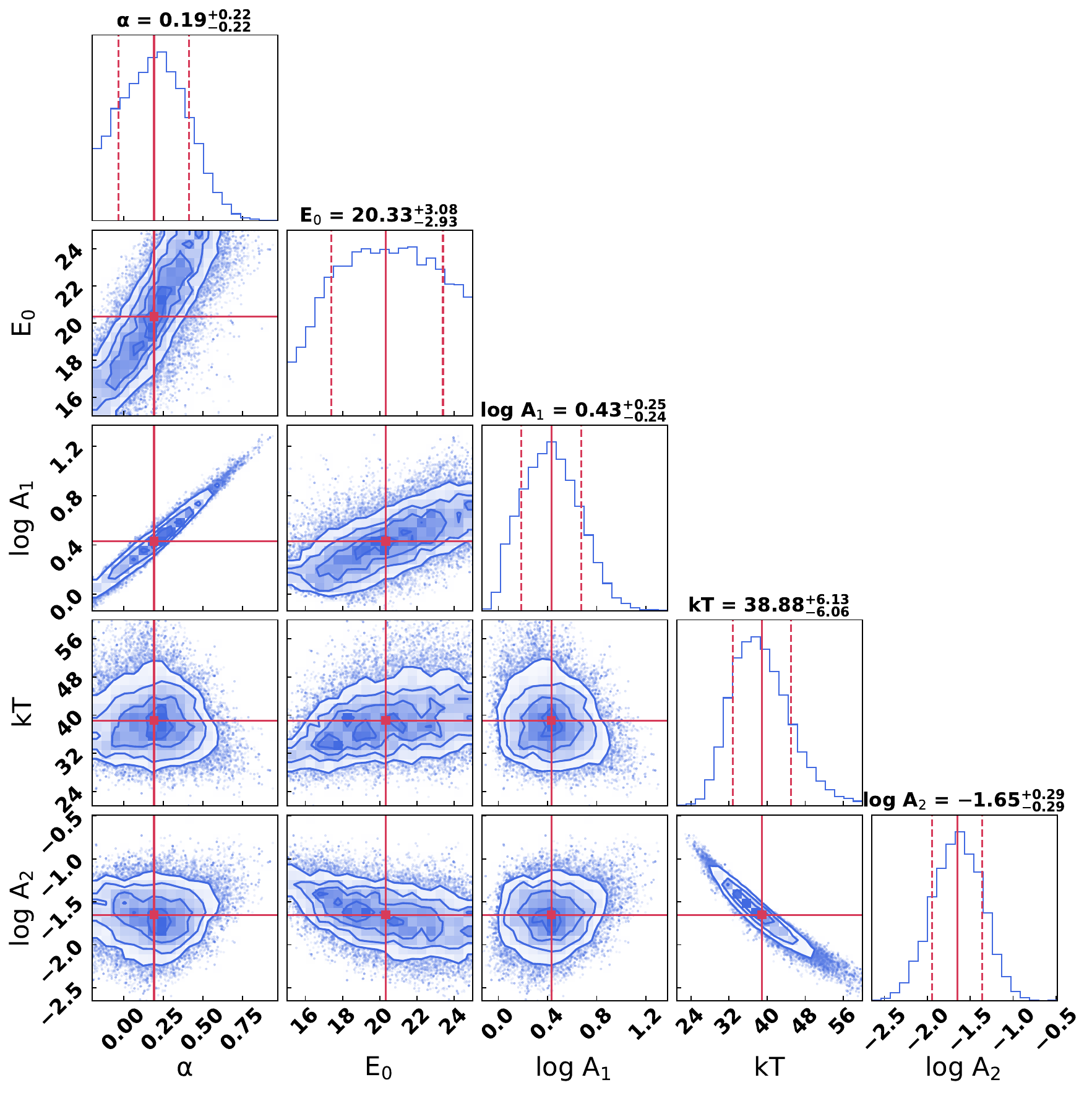}\\
	\includegraphics[angle=0,scale=0.40]{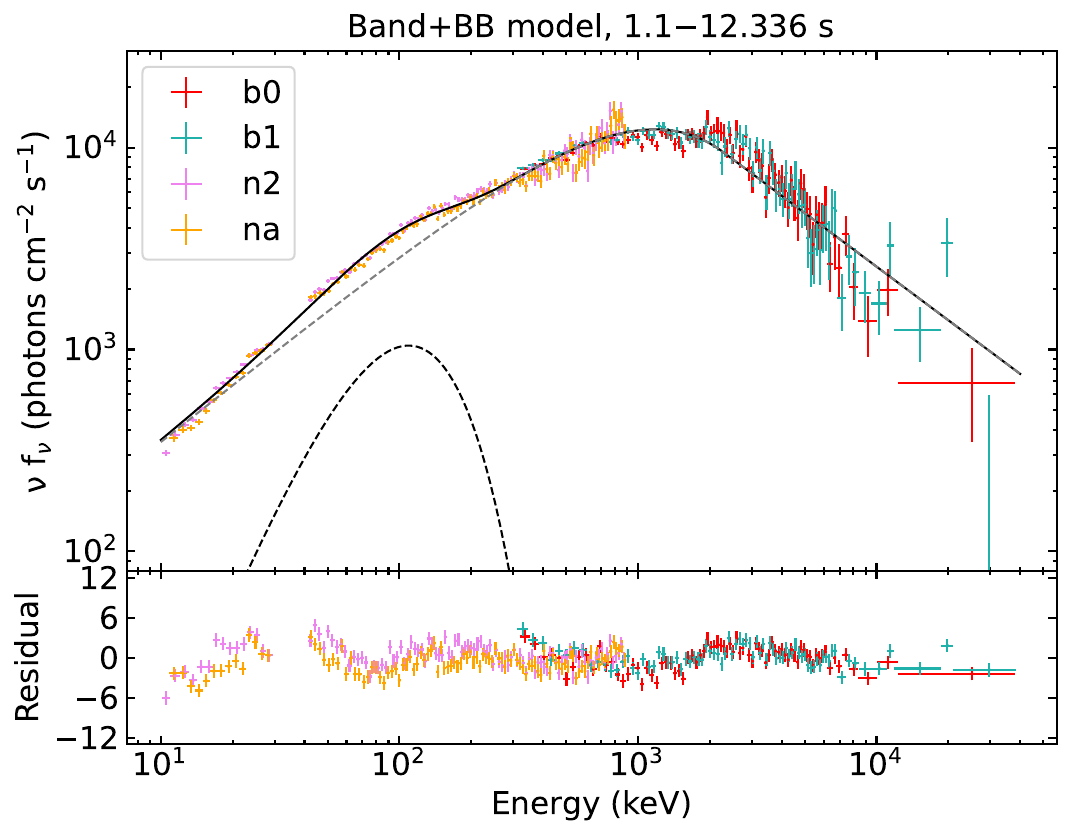}
	\includegraphics[angle=0,scale=0.16]{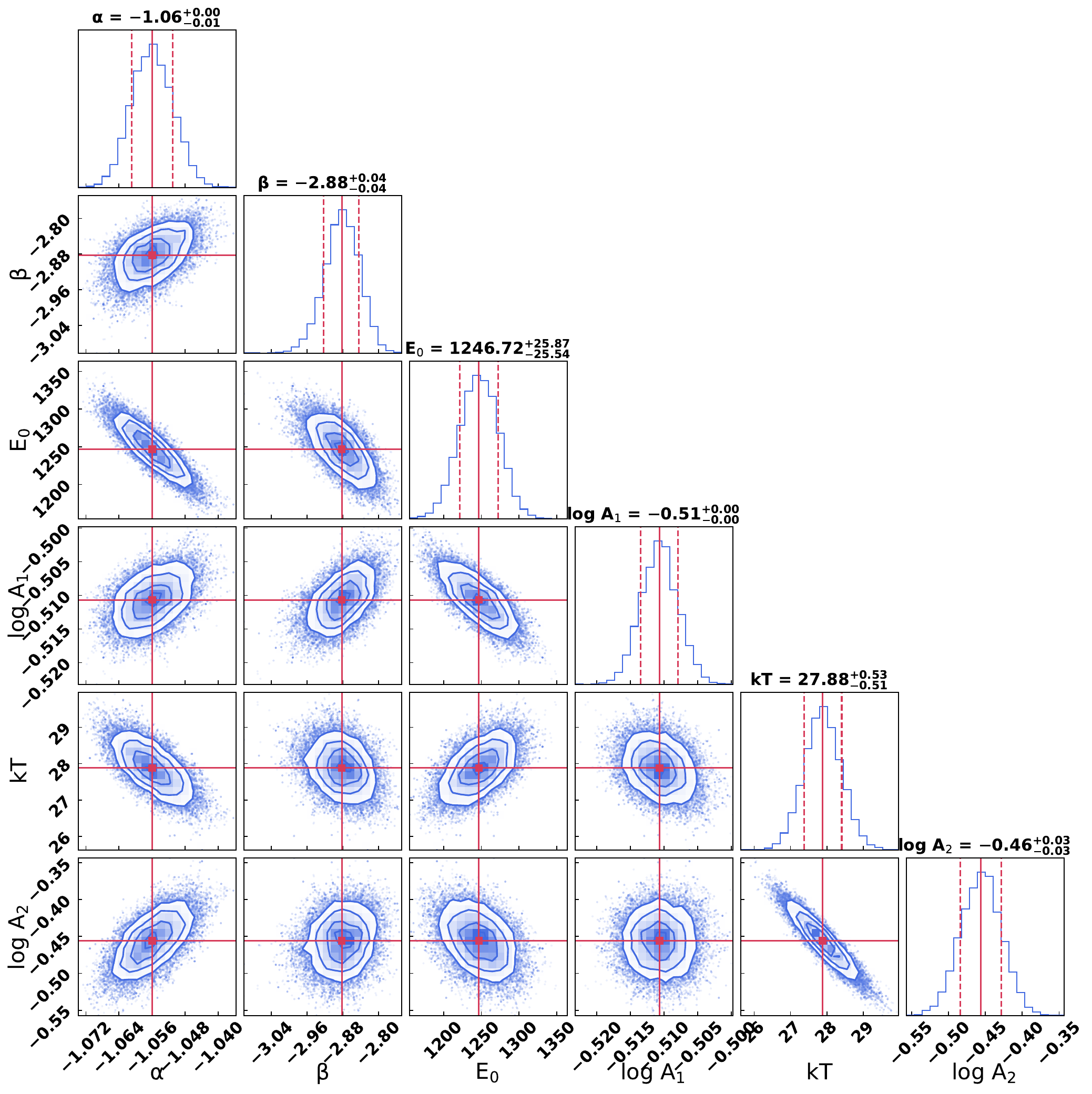}\\
	\includegraphics[angle=0,scale=0.40]{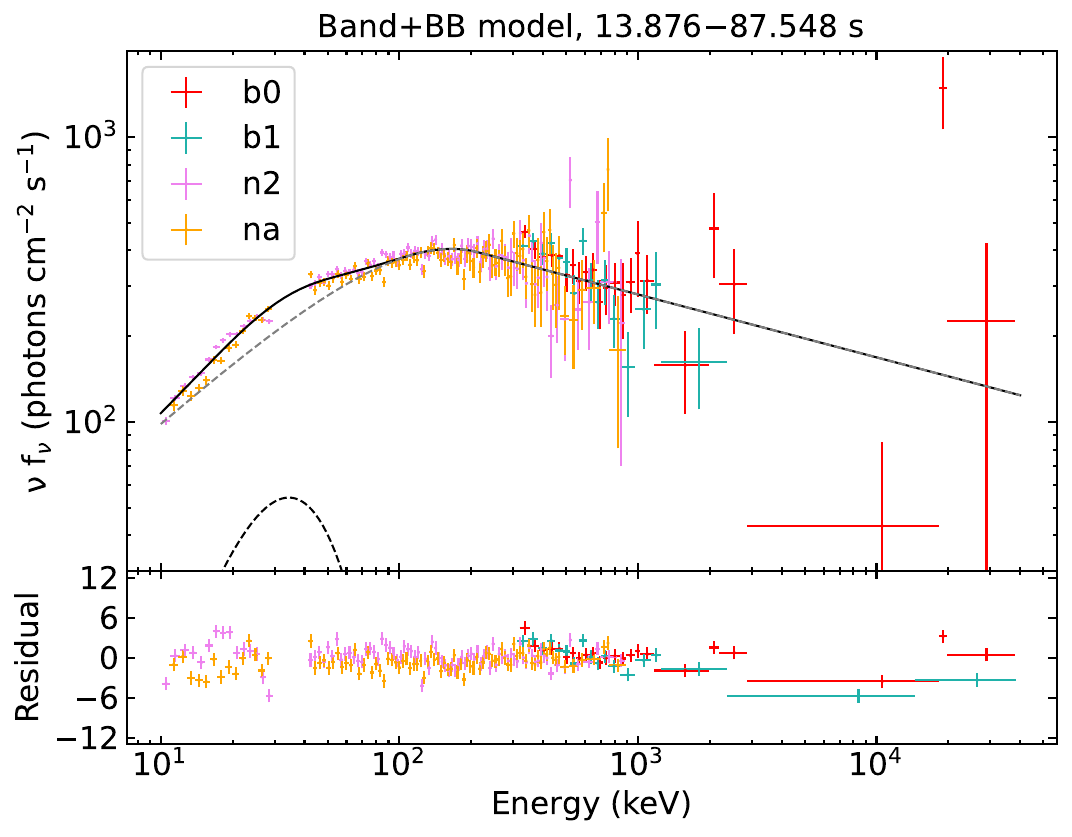}
	\includegraphics[angle=0,scale=0.16]{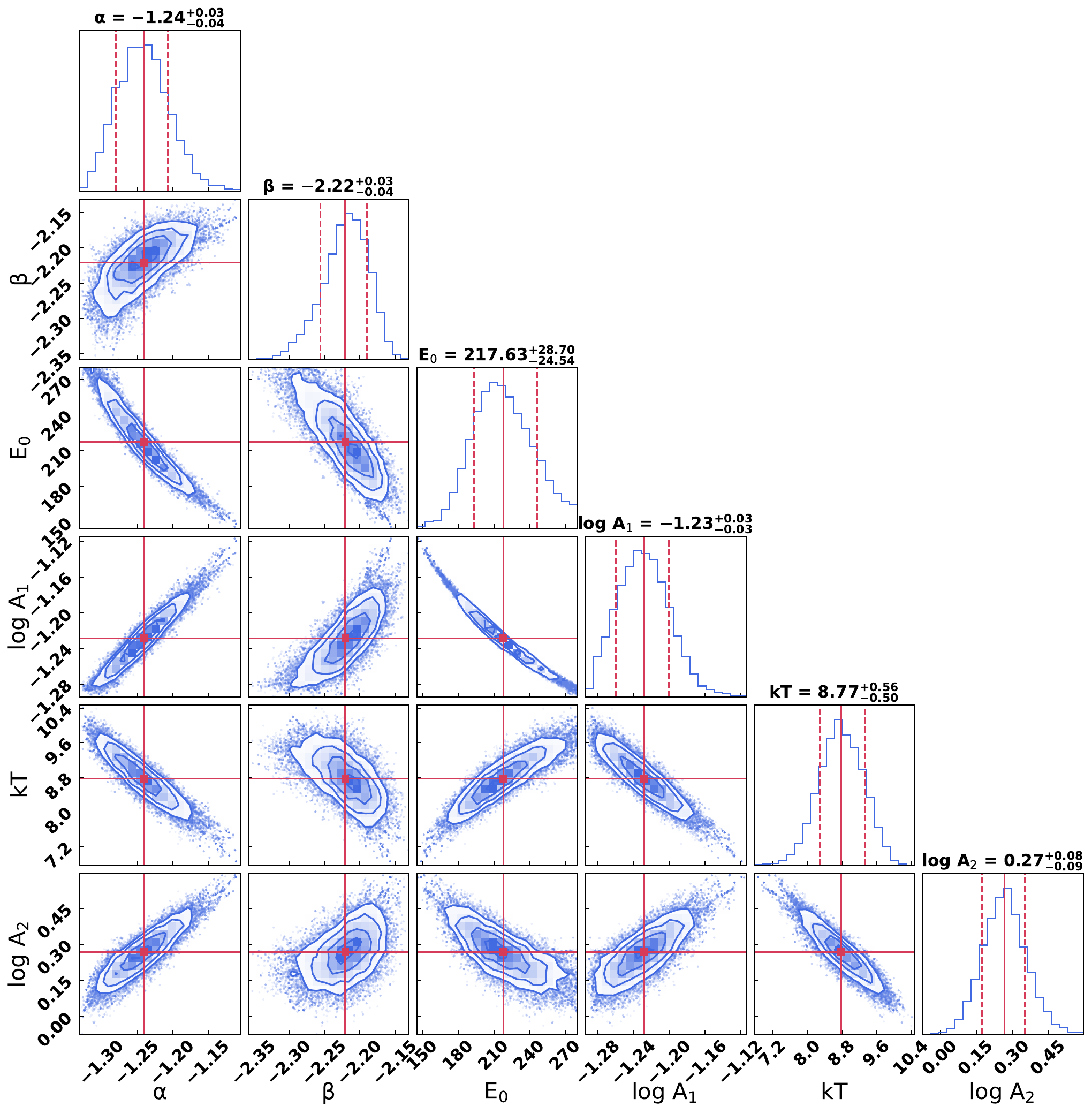}\\
	\includegraphics[angle=0,scale=0.40]{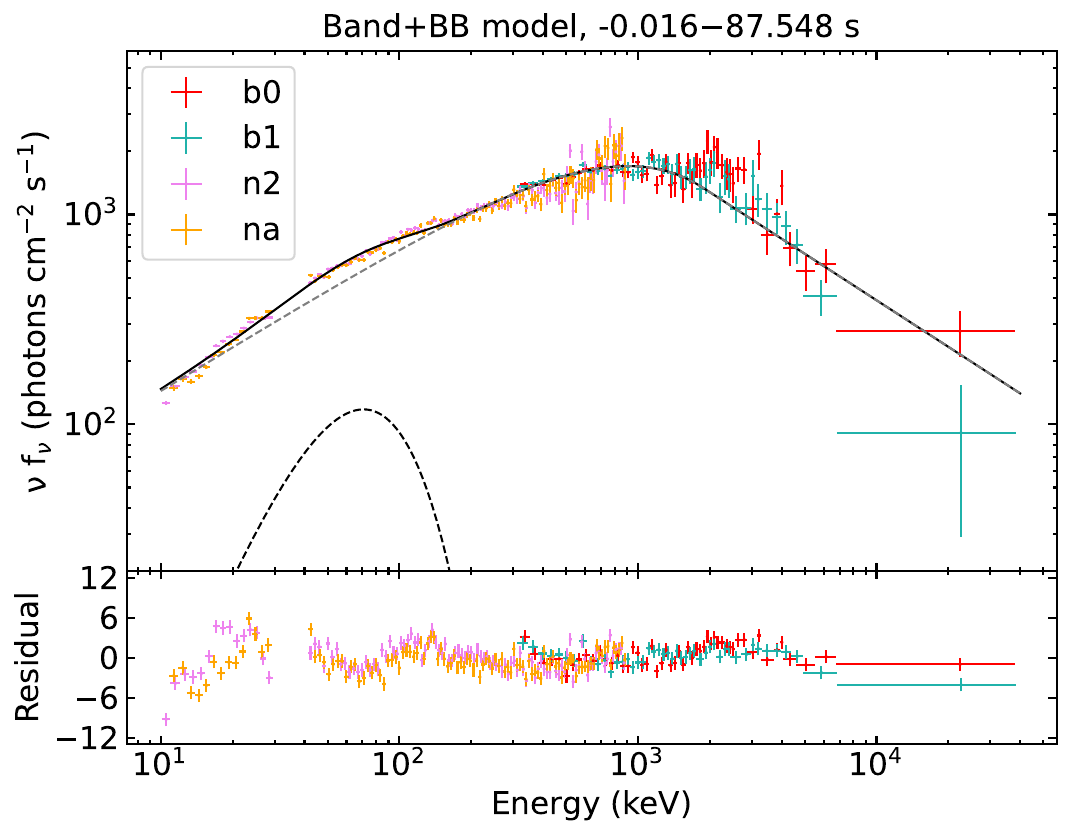}
	\includegraphics[angle=0,scale=0.16]{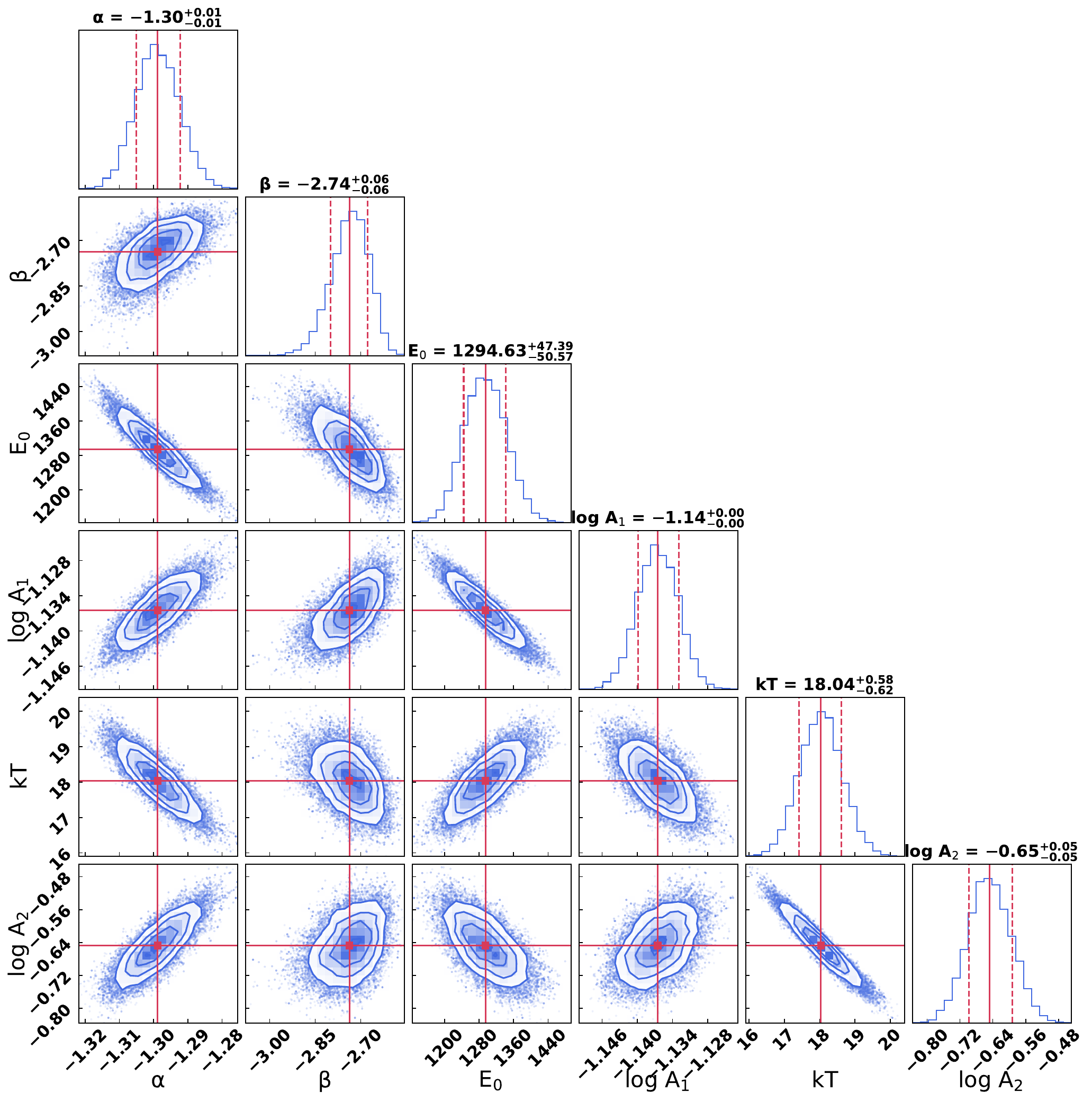}
	\caption{The spectra of GRB 211211A. The panels of each row from top to bottom represents the best-fit results for the precursor emission, main emission, extended emission, and whole emission episodes. The left panels show the spectrum of the best-fit model, and the right panels show the corner plot of the MCMC fitting results for the corresponding model.}
	\label{figure:specturm}
\end{figure*}

According to the standard fireball-shock model of GRBs, the observed spectrum is expected to be the superposition of two components: a non-thermal component from the internal shock \citep{1994ApJ...430L..93R,1997ApJ...490...92K} or Internal-Collision-induced MAgnetic Reconnection and Turbulence \citep[ICMART;][]{2011ApJ...726...90Z} in the optically thin region, and a quasi-thermal component from the fireball photosphere \citep{1994MNRAS.270..480T,2000ApJ...530..292M}.
Therefore, we employed five spectral models, including two non-thermal models: Band and CPL, as well as three combined models that incorporate both non-thermal and quasi-thermal components: Band+BB, CPL+BB, and PL+BB.
The spectral fitting results of the precursor emission, main emission, extended emission, and whole emission for each GRB are list in Table \ref{table:spectral_pe}, Table \ref{table:spectral_me}, Table \ref{table:spectral_ee}, and Table \ref{table:spectral_we}, respectively.
The spectra and the best-fit results of the precursor emission, main emission, and extended emission for GRB 211211A are shown in Figure \ref{figure:specturm}.

We find that 8 out of 27 GRBs have quasi-thermal components in their precursor emissions, accounting for 29.63\%.
Note that the identification of the quasi-thermal component of the precursor emission may be affected by their luminosities.
There are 18 out of 29 GRBs that have quasi-thermal components in their main emissions, accounting for 62.07\%, which is a higher proportion than that in the precursor emissions. 
This indicates that the energy spectrum of some GRBs changes from non-thermal to quasi-thermal.
Furthermore, 6 out of 24 GRBs have quasi-thermal components in their extended emissions, accounting for 25\%, and 17 out of 28 GRBs have quasi-thermal components in their whole emissions, accounting for 58.62\%.
There are four GRBs, GRB 110825A, GRB 200914A, GRB 211211A, and GRB 230307A, that exhibited quasi-thermal components in each episode and in the whole emission.
The $kT$ of GRB 110825A first increases and then decreases, whereas the $kT$ of GRB 211211A and GRB 230307A gradually decreases over time.
For GRB 200914A, the $kT$ of both the main emission and extended emission drops below the detection threshold of Fermi/GBM, making it impossible to determine its evolution.

\begin{table*}
	\caption{The spectral fitting results for precursor emissions of the three-episode GRBs}
	\label{table:spectral_pe}
	\scriptsize
	\setlength{\tabcolsep}{8pt}
	\renewcommand{\arraystretch}{1.3}
	\begin{tabular}{lcccccccccc}
		\hline
		GRB & Model & $\alpha$ & $\beta$ & $E_{\rm p}$ & $kT$ & Flux$_{\rm BB,-7}$  & Flux$_{\rm tot,-7}$  & pgstat/dof & BIC \\
		&  &  &  & (keV) & (keV) & (erg cm$^{-2}$ s$^{-1}$) & (erg cm$^{-2}$ s$^{-1}$) &  &  & \\
		\hline
		090131 & Band & $-1.36_{-0.12}^{+0.12}$ & $-3.42_{-1.05}^{+1.11}$ & $1707.19_{-950.91}^{+1073.06}$ & -- & -- & $6.68_{-1.21}^{+0.91}$ & 345.52/298 & 1.89 \\
		& CPL & $-1.35_{-0.13}^{+0.11}$ &  & $1769.41_{-1030.29}^{+1041.45}$ & -- & -- & $6.83_{-1.3}^{+1.36}$ & 345.52/299 & -0.34 \\
		& Band+BB & $-1.22_{-0.18}^{+0.37}$ & $-3.40_{-1.11}^{+1.07}$ & $1855.68_{-1492.06}^{+1259.22}$ & $4.24_{-1.78}^{+2.20}$ & $0.21_{-0.11}^{+0.11}$ & $7.29_{-1.21}^{+0.61}$ & 342.08/296 & 8.10 \\
		& CPL+BB & $-1.05_{-0.26}^{+0.28}$ & -- & $1031.17_{-641.54}^{+641.19}$ & $4.74_{-2.04}^{+1.94}$ & $0.56_{-0.11}^{+0.12}$ & $7.49_{-1.56}^{+0.76}$ & 342.02/297 & 5.22 \\
		& PL+BB & $-1.64_{-0.17}^{+0.16}$ & -- & -- & $109.27_{-53.75}^{+56.47}$ & $2.79_{-0.95}^{+0.97}$ & $5.53_{-0.32}^{+1.95}$ & 343.51/298 & -0.58 \\
		\hline
		090618 & Band & $-0.65_{-0.07}^{+0.06}$ & $-2.36_{-0.09}^{+0.09}$ & $161.38_{-17.42}^{+19.19}$ & -- & -- & $9.35_{-0.24}^{+0.09}$ & 478.01/191 & -37.04 \\
		& CPL & $-0.82_{-0.04}^{+0.04}$ &  & $203.96_{-15.44}^{+15.51}$ & -- & -- & $8.81_{-0.25}^{+0.18}$ & 511.09/192 & -17.92 \\
		& Band+BB & $-0.66_{-0.07}^{+0.08}$ & $-2.38_{-0.10}^{+0.11}$ & $166.36_{-23.89}^{+21.80}$ & $13.06_{-8.71}^{+9.58}$ & $0.05_{-0.05}^{+0.05}$ & $9.3_{-0.19}^{+0.22}$ & 476.82/189 & -28.09 \\
		& CPL+BB & $-1.09_{-0.08}^{+0.08}$ & -- & $320.39_{-76.34}^{+80.76}$ & $25.51_{-2.07}^{+2.20}$ & $1.27_{-0.06}^{+0.07}$ & $9.62_{-0.52}^{+0.01}$ & 489.50/190 & -24.44 \\
		& PL+BB & $-1.65_{-0.02}^{+0.02}$ & -- & -- & $32.69_{-0.95}^{+0.95}$ & $3.27_{-0.08}^{+0.08}$ & $8.63_{-0.14}^{+0.16}$ & 575.63/191 & 54.27 \\
		\hline
		\multicolumn{11}{l}{This table is published in its entirety in the machine-readable format. A full machine-readable version is available online.}\\
	\end{tabular}
\end{table*}

\begin{table*}
	\caption{The spectral fitting results for main emissions of the three-episode GRBs}
	\label{table:spectral_me}
	\scriptsize
	\setlength{\tabcolsep}{8pt}
	\renewcommand{\arraystretch}{1.3}
	\begin{tabular}{lcccccccccc}
		\hline
		GRB & Model & $\alpha$ & $\beta$ & $E_{\rm p}$ & $kT$ & Flux$_{\rm BB,-7}$  & Flux$_{\rm tot,-7}$  & pgstat/dof & BIC \\
		&  &  &  & (keV) & (keV) & (erg cm$^{-2}$ s$^{-1}$) & (erg cm$^{-2}$ s$^{-1}$) &  &  & \\
		\hline
		090131 & Band & $-0.94_{-0.13}^{+0.14}$ & $-2.80_{-0.16}^{+0.16}$ & $48.67_{-10.32}^{+10.15}$ & -- & -- & $9.13_{-0.46}^{+0.17}$ & 487.94/333 & 296.34 \\
		& CPL & $-1.17_{-0.06}^{+0.05}$ & -- & $53.87_{-5.65}^{+5.63}$ & -- & -- & $8.05_{-0.19}^{+0.09}$& 506.79/334 & 295.33 \\
		& Band+BB & $-0.87_{-0.19}^{+0.21}$ & $-3.01_{-0.20}^{+0.21}$ & $62.00_{-15.25}^{+15.64}$ & $6.17_{-0.83}^{+0.80}$ & $0.89_{-0.05}^{+0.05}$ & $8.89_{-0.44}^{+0.03}$ & 474.79/331 & 289.63 \\
		& CPL+BB & $-1.12_{-0.14}^{+0.14}$ & -- & $68.15_{-15.79}^{+16.26}$ & $7.16_{-0.89}^{+0.95}$ & $1.04_{-0.06}^{+0.06}$ & $8.39_{-0.35}^{+0.004}$ & 485.06/332 & 288.60 \\
		\hline
		090618 & Band & $-1.03_{-0.01}^{+0.01}$ & $-2.39_{-0.03}^{+0.03}$ & $172.84_{-6.86}^{+6.77}$ & -- & -- & $40.8_{-0.28}^{+0.18}$ & 854.69/225 & 129.03 \\
		& CPL & $-1.15_{-0.01}^{+0.01}$ & -- & $224.99_{-5.49}^{+5.45}$ & -- & -- & $40.7_{-0.26}^{+0.25}$ & 1044.05/226 & 291.95 \\
		& Band+BB & $-0.91_{-0.04}^{+0.05}$ & $-2.38_{-0.03}^{+0.03}$ & $170.16_{-12.99}^{+12.14}$ & $7.02_{-0.54}^{+0.51}$ & $0.86_{-0.04}^{+0.04}$ & $40.8_{-0.31}^{+0.17}$ & 832.48/223 & 118.49 \\
		& CPL+BB & $-1.32_{-0.02}^{+0.02}$ & -- & $298.95_{-19.08}^{+19.46}$ & $25.90_{-0.76}^{+0.78}$ & $4.13_{-0.09}^{+0.10}$ & $41.5_{-0.33}^{+0.16}$ & 881.44/224 & 145.55 \\
		\hline
		\multicolumn{11}{l}{This table is published in its entirety in the machine-readable format. A full machine-readable version is available online.}\\
	\end{tabular}
\end{table*}

\begin{table*}
	\caption{The spectral fitting results for extended emissions of the three-episode GRBs}
	\label{table:spectral_ee}
	\scriptsize
	\setlength{\tabcolsep}{8pt}
	\renewcommand{\arraystretch}{1.3}
	\begin{tabular}{lcccccccccc}
		\hline
		GRB & Model & $\alpha$ & $\beta$ & $E_{\rm p}$ & $kT$ & Flux$_{\rm BB,-7}$  & Flux$_{\rm tot,-7}$  & pgstat/dof & BIC \\
		&  &  &  & (keV) & (keV) & (erg cm$^{-2}$ s$^{-1}$) & (erg cm$^{-2}$ s$^{-1}$) &  &  & \\
		\hline
		090131 & Band & $-1.49_{-0.08}^{+0.07}$ & $-2.51_{-0.32}^{+0.28}$ & $116.92_{-37.49}^{+41.82}$ & -- & -- & $2.10_{-0.12}^{+0.07}$ & 689.09/333 & -1.8 \\
		& CPL & $-1.53_{-0.06}^{+0.05}$ & -- & $131.28_{-32.78}^{+33.48}$ & -- & -- & $2.02_{-0.13}^{+0.06}$ & 690.14/334 & -8.09 \\
		& Band+BB & $-1.53_{-0.07}^{+0.08}$ & $-2.26_{-0.14}^{+0.16}$ & $108.60_{-37.07}^{+39.38}$ & $18.14_{-5.30}^{+5.94}$ & $0.05_{-0.02}^{+0.02}$ & $2.10_{-0.11}^{+0.09}$ & 688.38/331 & 3.93 \\
		& CPL+BB & $-1.56_{-0.07}^{+0.07}$ & -- & $139.01_{-42.85}^{+45.02}$ & $15.84_{-7.19}^{+6.92}$ & $0.03_{-0.02}^{+0.02}$ & $2.02_{-0.10}^{+0.07}$ & 687.83/332 & -0.69 \\
		\hline
		090618 & Band & $-1.31_{-0.06}^{+0.06}$ & $-3.81_{-0.67}^{+0.60}$ & $46.39_{-5.88}^{+5.86}$ & -- & -- & $2.87_{-0.06}^{+0.10}$ & 1427.95/225 & 99.41 \\
		& CPL & $-1.34_{-0.06}^{+0.06}$ & -- & $46.87_{-5.72}^{+5.81}$ & -- & -- & $2.81_{-0.05}^{+0.05}$ & 1430.30/226 & 93.47 \\
		& Band+BB & $-1.34_{-0.08}^{+0.08}$ & $-3.46_{-0.36}^{+0.33}$ & $48.64_{-9.03}^{+9.31}$ & $7.19_{-2.05}^{+1.92}$ & $0.09_{-0.02}^{+0.02}$ & $2.92_{-0.06}^{+0.07}$ & 1425.63/223 & 107.53 \\
		& CPL+BB & $-1.36_{-0.08}^{+0.07}$ & -- & $48.38_{-8.45}^{+8.55}$ & $7.40_{-3.87}^{+3.36}$ & $0.06_{-0.02}^{+0.02}$ & $2.81_{-0.04}^{+0.07}$ & 1427.14/224 & 102.44 \\
		\hline
		\multicolumn{11}{l}{This table is published in its entirety in the machine-readable format. A full machine-readable version is available online.}\\
	\end{tabular}
\end{table*}

\begin{table*}
	\caption{The spectral fitting results for whole emissions of the three-episode GRBs}
	\label{table:spectral_we}
	\scriptsize
	\setlength{\tabcolsep}{8pt}
	\renewcommand{\arraystretch}{1.3}
	\begin{tabular}{lcccccccccc}
		\hline
		GRB & Model & $\alpha$ & $\beta$ & $E_{\rm p}$ & $kT$ & Flux$_{\rm BB,-7}$  & Flux$_{\rm tot,-7}$  & pgstat/dof & BIC \\
		&  &  &  & (keV) & (keV) & (erg cm$^{-2}$ s$^{-1}$) & (erg cm$^{-2}$ s$^{-1}$) &  &  & \\
		\hline
		090131 & Band & $-1.28_{-0.09}^{+0.08}$ & $-2.38_{-0.10}^{+0.10}$ & $65.26_{-14.42}^{+14.32}$ & -- & -- & $2.48_{-0.11}^{+0.03}$ & 914.35/333 & 56.03 \\
		& CPL & $-1.46_{-0.05}^{+0.04}$ & -- & $81.31_{-12.19}^{+12.35}$ & -- & -- & $2.17_{-0.08}^{+0.03}$ & 929.72/334 & 57.37 \\
		& Band+BB & $-1.27_{-0.10}^{+0.11}$ & $-2.44_{-0.11}^{+0.13}$ & $72.40_{-19.37}^{+16.49}$ & $5.65_{-2.11}^{+2.30}$ & $0.04_{-0.01}^{+0.01}$ & $2.44_{-0.10}^{+0.04}$ & 910.98/331 & 65.87 \\
		& CPL+BB & $-1.52_{-0.07}^{+0.08}$ & -- & $101.53_{-28.83}^{+29.86}$ & $9.91_{-2.03}^{+1.85}$ & $0.14_{-0.02}^{+0.02}$ & $2.30_{-0.13}^{+0.05}$ & 920.13/332 & 60.3 \\
		\hline
		090618 & Band & $-1.14_{-0.01}^{+0.01}$ & $-2.36_{-0.03}^{+0.03}$ & $145.49_{-6.10}^{+6.29}$ & -- & -- & $13.65_{-0.09}^{+0.09}$ & 2239.26/225 & 421.67 \\
		& CPL & $-1.24_{-0.01}^{+0.01}$ & -- & $186.14_{-5.71}^{+5.56}$ & -- & -- & $13.28_{-0.11}^{+0.09}$ & 2389.09/226 & 548.64 \\
		& Band+BB & $-1.04_{-0.04}^{+0.04}$ & $-2.37_{-0.03}^{+0.03}$ & $146.68_{-10.69}^{+11.19}$ & $6.82_{-0.56}^{+0.59}$ & $0.28_{-0.02}^{+0.02}$ & $13.67_{-0.09}^{+0.06}$ & 2219.55/223 & 413.99 \\
		& CPL+BB & $-1.39_{-0.02}^{+0.02}$ & -- & $239.30_{-21.06}^{+19.59}$ & $22.73_{-1.04}^{+0.96}$ & $1.05_{-0.03}^{+0.03}$ & $13.75_{-0.15}^{+0.06}$ & 2308.22/224 & 485.6 \\
		\hline
		\multicolumn{11}{l}{This table is published in its entirety in the machine-readable format. A full machine-readable version is available online.}\\
	\end{tabular}
\end{table*}

In this work, $E_{\rm iso}$ are corrected to that in the energy band of 1--$10^{4}$ keV in the rest frame.
$E_{\rm iso} = \frac{4\pi D_{\rm L}^2 S_{\gamma}k}{(1+z)}$, where $D_{\rm L}$ is the luminosity distance, $S_{\gamma}$ is the fluence, $z$ is the redshift, and $k$ is the $k$--correction factor \citep{2007ApJ...660...16S}.
We assume a flat universe ($\Omega_{\rm k} = 0$) with the cosmological parameters $H_{0}=67.3$ km s$^{-1}$ Mpc$^{-1}$, $\Omega_{\rm M}=0.315$, and $\Omega_{\rm \Lambda}=0.685$.
The peak energy in the rest frame is calculated as $E_{\rm p,z} = E_{\rm p}(1 + z)$.

\begin{figure*}
	\centering
	\includegraphics[angle=0,scale=0.39]{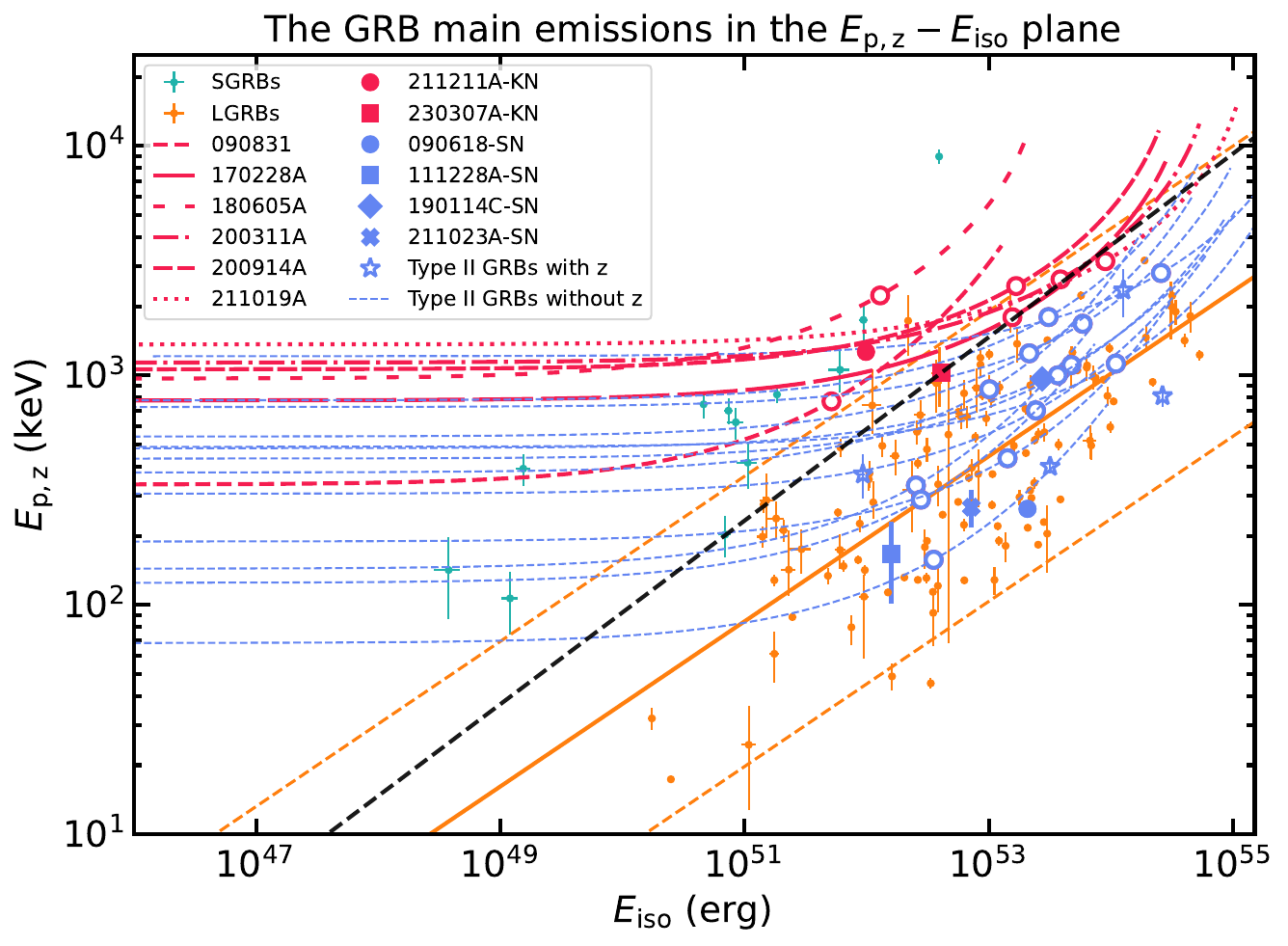}
	\includegraphics[angle=0,scale=0.39]{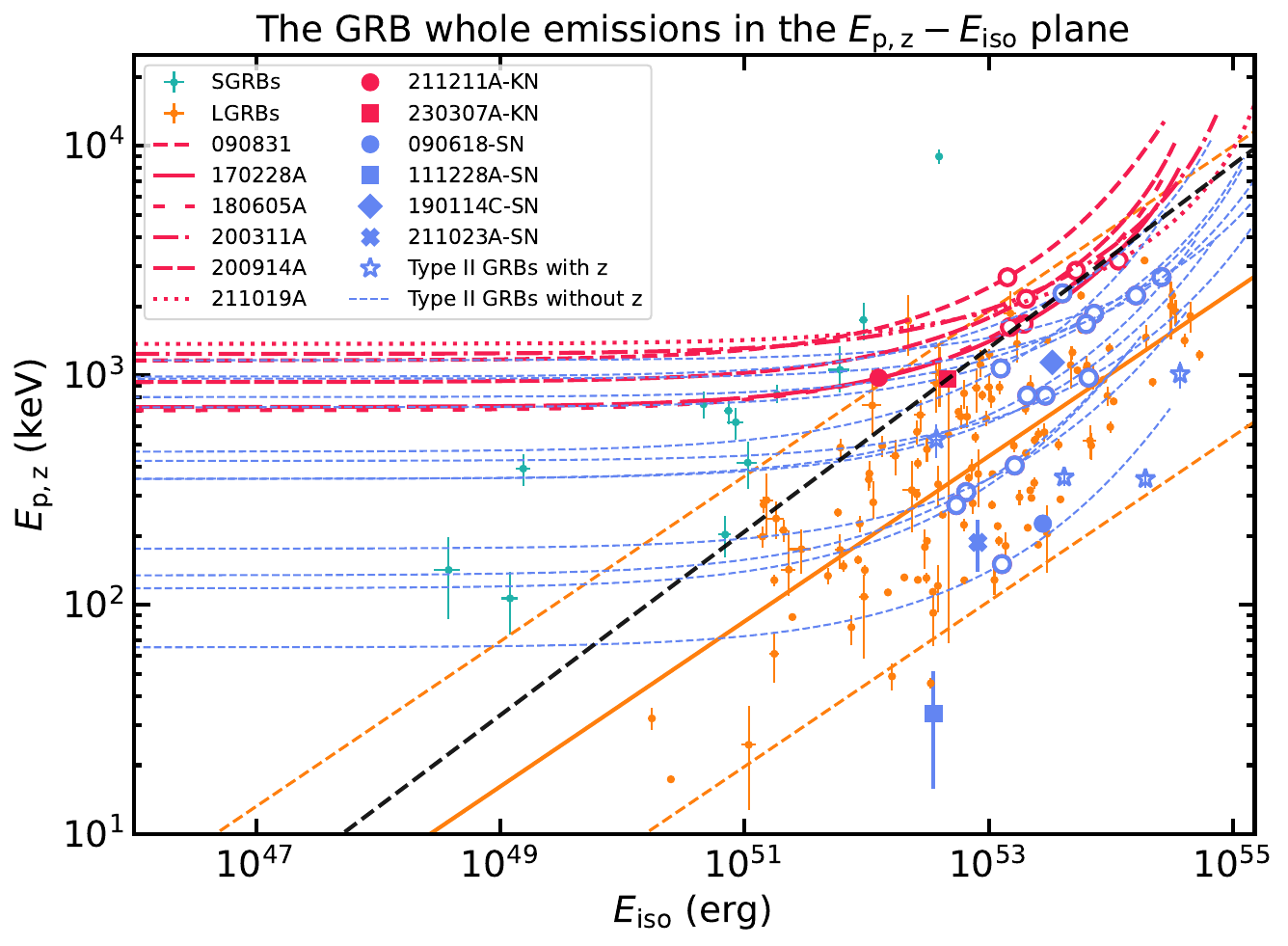}\\
	\caption{The $E_{\rm p,z}$--$E_{\rm iso}$ planes for different episodes of GRBs. The solid red and blue markers represent GRBs associated with KNe and SNe, respectively, where different markers denote different GRBs. The red and blue dashed lines are GRBs classified as Type IL and Type II by the machine learning, respectively. The red and blue hollow circles are GRBs without redshift at $z = 1.31$. The orange and green circles are LGRBs and SGRBs, respectively. The orange solid line is the best-fit line for LGRBs, $E_{\rm p,z} \propto E_{\rm iso}^{0.36}$. The orange dashed lines are the 2$\sigma$ confidence region of LGRBs. The black dashed line is the boundary with a slope of 0.4 passing through GRB 230307A.}
	\label{figure:Ep-Eiso}
\end{figure*}

We calculate the energy of each episode for GRBs with redshift, a total of 10 GRBs, and analyze their properties in the $E_{\rm p,z}$--$E_{\rm iso}$ plane. 
For GRBs without redshift, we simulate their energy across redshifts from 0.0001 to 10, and the results are shown in Figure \ref{figure:Ep-Eiso}.
Obviously, the main emissions of both GRB 090831, GRB 180605A, and GRB 200914A deviate from Type II GRBs regardless of their redshift, suggesting that they may be Type IL GRBs.
For two Type IL GRBs, the main and whole emissions of GRB 211211A deviate from Type II GRBs, while GRB 230307A does not.

To quantitatively analyze GRBs without known redshifts, we adopt the energy-hardness ($EH$) parameter as a criterion, $EH = (E_{\rm p,z}/100)/(E_{\rm iso}/10^{51})^{0.4}$ \citep{2020MNRAS.492.1919M}, and use the $EH$ of GRB 230307A as the boundary.
For main emission, we find that GRB 090831, GRB 170228A, GRB 180605A, GRB 200311A, and GRB 200914A never cross the boundary when $z < 1.31$, which is the typical value of GRBs with $z$, indicating that the probability of $z > 1.31$ is 50\%.
GRB 211019A crosses the boundary within the $z$ range of 0.83--4.32, with a probability of 58\% based on the cumulative redshift distribution function \citep{2024ApJ...976...62Z}. 
Although this probability is slightly greater than 50\%, it depends on $EH$, which is sensitive to the slope of the $E_{\rm p,z}$--$E_{\rm iso}$ correlation. 
Moreover, even if it crosses the boundary, the deviation is very small.
Furthermore, all other GRBs significantly cross the boundary, indicating that they are most likely Type II GRBs. 
The overall trend for the whole emission is highly similar to that of the main emission.

\section{New Insight for Classification based on Machine Learning} \label{sec:classification}

We find that GRBs with the three-episode structure can be produced by either mergers or collapsars, while all of them are classified as LGRBs.
GRB 211211A and GRB 230307A are confirmed Type IL GRBs, while GRB 090618A, GRB 111228A, GRB 190114C, and GRB 211023A are confirmed Type II GRBs, all of which can be divided into three episodes.
Although some clues were found in the $E_{\rm p,z}$--$E_{\rm iso}$ plane, we do not find significant differences in their temporal and spectral properties.
Therefore, we apply machine learning, which has been widely used in GRB classification, to distinguish them.

\begin{figure*}
	\centering
	\includegraphics[angle=0,scale=0.55]{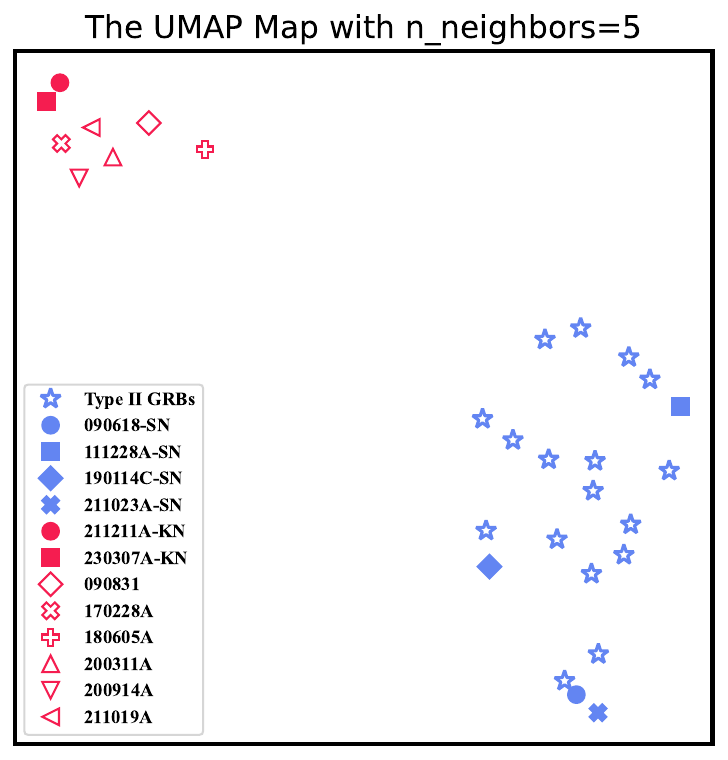}
	\includegraphics[angle=0,scale=0.55]{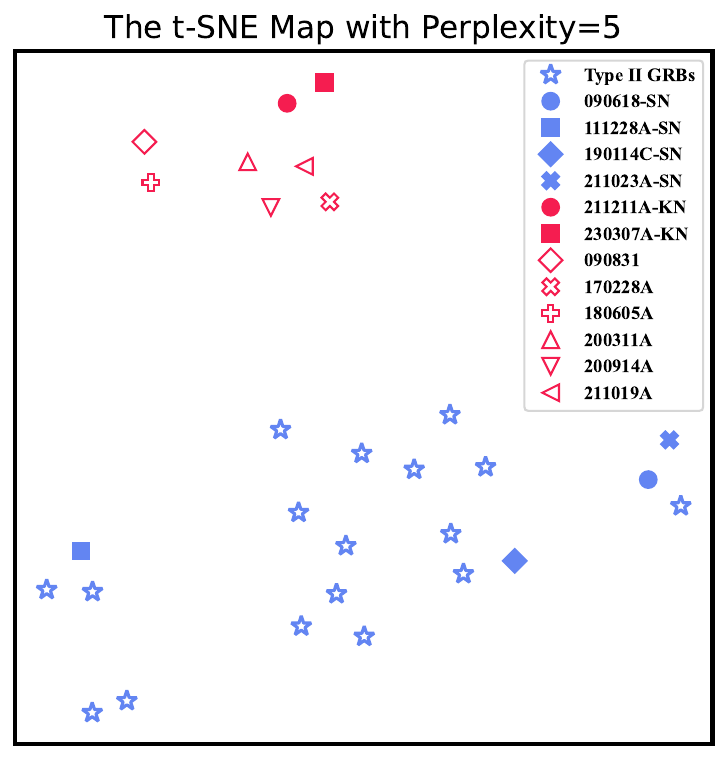}
	\includegraphics[angle=0,scale=0.55]{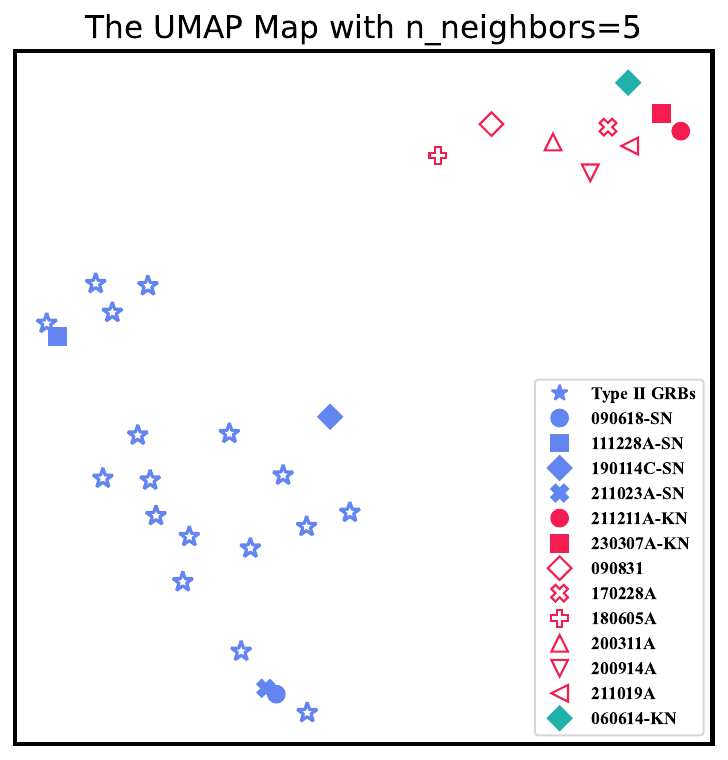}
	\includegraphics[angle=0,scale=0.55]{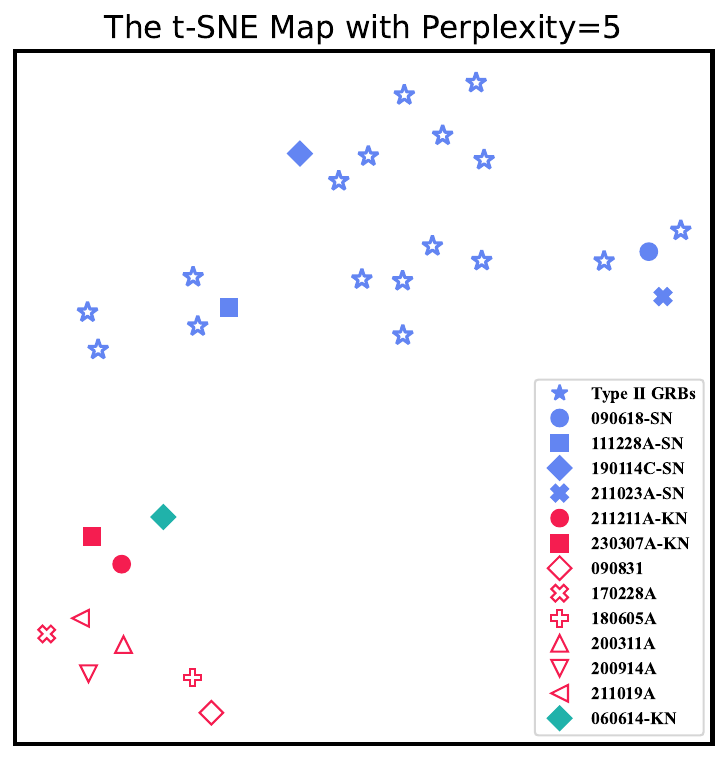}
	\caption{The UMAP and t-SNE maps. The red and blue markers represent the two clusters classified by GMM. The hollow red and blue markers represent Type IL and Type II GRBs identified by machine learning without electromagnetic counterparts. Different hollow red markers represent different Type IL GRBs without electromagnetic counterparts, while all Type II GRBs without electromagnetic counterparts are uniformly represented by hollow star markers. The green marker is GRB 060614, which is classified as a Type IL GRB.}
	\label{figure:classification}
\end{figure*}

Based on 12 parameters, $T_{\rm 100,PE}$, $T_{\rm 100,ME}$, $T_{\rm 100,EE}$, $T_{\rm 100,WE}$, $T_{\rm wt1}$, $T_{\rm wt2}$, $T_{\rm MVT,PE}$, $T_{\rm MVT,ME}$, $\tau_{\rm 42,ME}$, $\tau_{\rm 42,WE}$, $E_{\rm p,ME}$, and $E_{\rm p,WE}$, we applie t-SNE and UMAP to the three-episode GRBs.
Note that, the catalog was a logarithmic transformation (except for $\tau_{\rm 42,ME}$ and $\tau_{\rm 42,WE}$ because they contain negative values) and standardization before applying t-SNE and UMAP.
The most critical parameter determining the t-SNE result is $perplexity$, which we set to $perplexity = 5$. 
For the UMAP result, the key parameters are $n\_neighbors$ and $min\_dist$, with $n\_neighbors = 5$ and $min\_dist = 0.001$.
The results are shown in Figure \ref{figure:classification}.

Visually, GRBs are distinctly divided into two clusters.
To further validate this classification, we apply the Gaussian Mixture Model (GMM) method to the t-SNE and UMAP results and calculate the BIC for one, two, and three cluster cases to determine the optimal number of clusters, respectively.
The BIC results indicate that the two-cluster classification is optimal in all cases.

\begin{figure*}
	\centering
	\includegraphics[angle=0,scale=0.23]{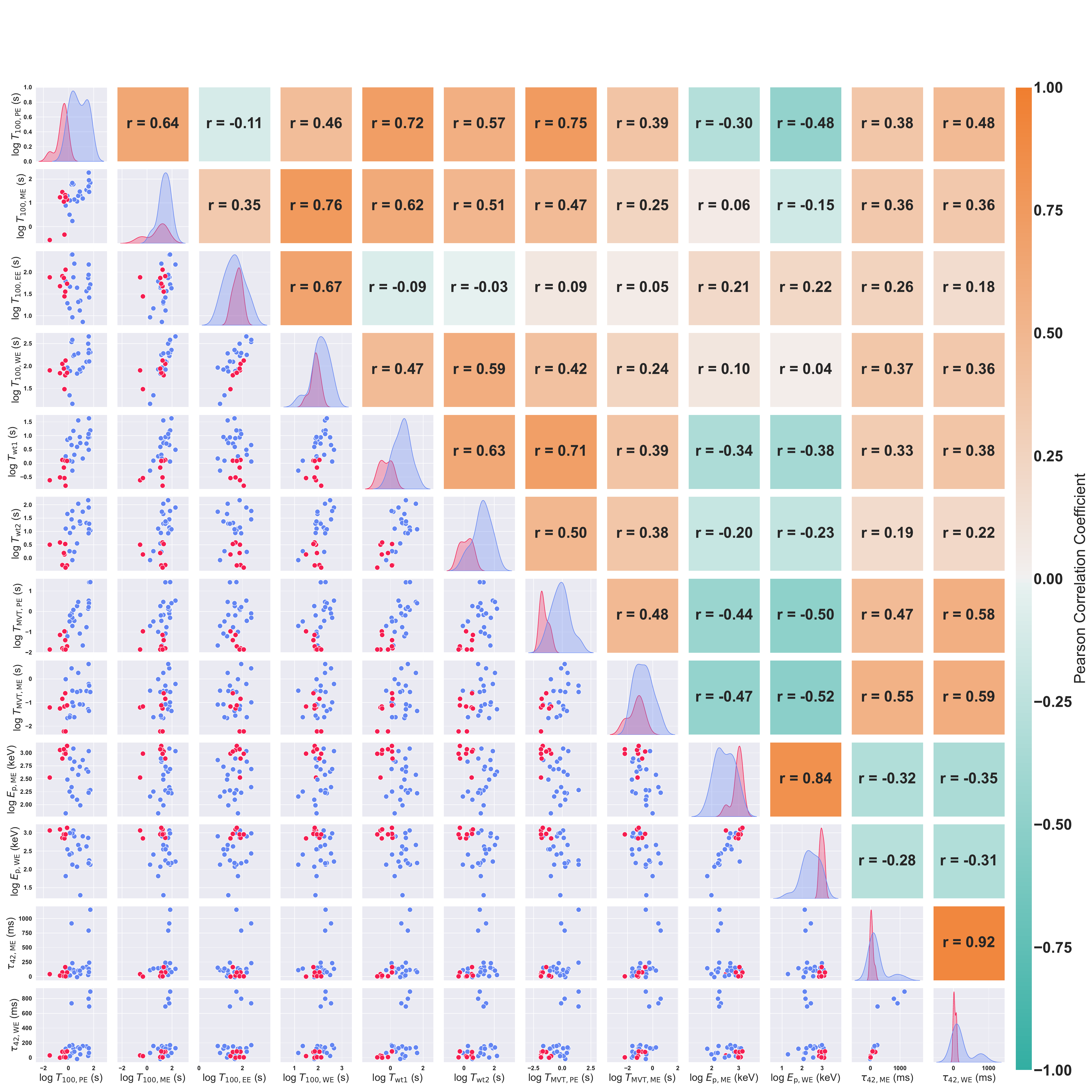}
	\caption{The comprehensive statistical plot of 12 parameters applied for machine learning. The lower-left corner displays scatter plots between pairs of parameters. The diagonal line displays the kernel density plots of the parameters. The upper-right corner displays the Pearson correlation coefficients between the parameters without classification, where orange and green colors represent the tight positive correlations and negative correlations, respectively. The red and blue markers represent Type IL and Type II GRBs identified by machine learning, respectively.}
	\label{figure:par}
\end{figure*}

Interestingly, Type IL GRBs and Type II GRBs are classified into different clusters, suggesting that the two clusters may correspond to different progenitors.
Among them, eight GRBs (GRB 090831, GRB 170228A, GRB 180605A, GRB 200311A, GRB 200914A, GRB 211019A, GRB 211211A, and GRB 230307A) form a separate cluster, potentially representing a population of LGRBs originated from merger, Type IL GRBs.
The remaining GRBs form another cluster, possibly corresponding to Type II GRBs.
To further verify this result, we plotted a comprehensive statistical plot of the 12 parameters, as shown in Figure \ref{figure:par}, and we will examine the properties of each Type IL GRB in detail. 
Furthermore, we note that the Type IL GRBs proposed by \cite{2025ApJ...979...73W} include three GRBs (GRB 170228A, GRB 211211A, and GRB 230307A), all of which are in our sample and belong to the Type IL GRBs.

GRB 090831 consists of a 0.03 s precursor emission, 0.28 s main emission, and 76.28 s extended emission, resulting in a total duration of 80 s. 
Its $T_{\rm wt1} = 0.24$ s, meaning the duration from the precursor emission onset to the end of the main emission is only 0.55 s, classifying it as an SGRB with EE. 
The spectral lags of the main and whole emissions are $\tau_{\rm 42,ME} = 41.71 \pm 3.02$ ms and $\tau_{\rm 42,WE} = 30.52 \pm 1.99$ ms, respectively.
The MVT of the whole emission occurs in the precursor emission, measuring 14 ms, and $T_{\rm MVT,ME} = 62$ ms.

GRB 170228A consists of a 0.32 s precursor emission, 28.51 s main emission, and 81.18 s extended emission, resulting in a total duration of 111.87 s, with $T_{\rm wt1} = 1.31$ s.
The spectral lags of the main and whole emissions are $\tau_{\rm 42,ME} = 72.65 \pm 4.12$ ms and $\tau_{\rm 42,WE} = 80.15 \pm 10.47$ ms, respectively.
The MVT of the whole emission occurs in the precursor emission, measuring 16 ms, and $T_{\rm MVT,ME} = 144$ ms. 

GRB 180605A consists of a 0.47 s precursor emission, 0.46 s main emission, and 27.85 s extended emission, resulting in a total duration of 30.43 s, with $T_{\rm wt1} = 0.29$ s. 
The spectral lags of the main and whole emissions are $\tau_{\rm 42,ME} = -1.33 \pm 1.36$ ms and $\tau_{\rm 42,WE} = 21.25 \pm 1.27$ ms, respectively.
The MVT of the whole emission occurs in the main emission, measuring 68 ms, and $T_{\rm MVT,PE} = 108$ ms.

GRB 200311A consists of a 0.21 s precursor emission, 17.14 s main emission, and 47.73 s extended emission, resulting in a total duration of 69.22 s, with $T_{\rm wt1} = 0.3$ s. 
The spectral lags of the main and whole emissions are $\tau_{\rm 42,ME} = 15.84 \pm 1.97$ ms and $\tau_{\rm 42,WE} = -0.73 \pm 2.96$ ms, respectively.
The MVT of the whole emission occurs in the main emission, measuring 56 ms, and $T_{\rm MVT,PE} = 72$ ms.

GRB 200914A consists of a 0.54 s precursor emission, 21.25 s main emission, and 35.76 s extended emission, resulting in a total duration of 62.14 s, with $T_{\rm wt1} = 1.23$ s. 
The spectral lags of the main and whole emissions are $\tau_{\rm 42,ME} = 162.63 \pm 48.37$ ms and $\tau_{\rm 42,WE} = 76.39 \pm 22.53$ ms, respectively.
The MVT of the whole emission occurs in the precursor emission, measuring 40 ms, and $T_{\rm MVT,ME} = 248$ ms.

GRB 211019A consists of a 0.76 s precursor emission, 13.59 s main emission, and 54.44 s extended emission, resulting in a total duration of 70.54 s, with $T_{\rm wt1} = 1.21$ s. 
The spectral lags of the main and whole emissions are $\tau_{\rm 42,ME} = 70.52 \pm 5.03$ ms and $\tau_{\rm 42,WE} = 84.89 \pm 1.01$ ms, respectively.
The MVT of the whole emission occurs in the precursor emission, measuring 20 ms, and $T_{\rm MVT,ME} = 76$ ms.

The temporal properties of the six Type IL GRB candidates are all consistent with those of Type I GRBs.
In addition, although their redshifts are unknown, the properties of main emissions are consistent with those of Type I GRBs in the $E_{\rm p,z}$--$E_{\rm iso}$ plane.
These properties support that they are also Type IL GRBs, similar to GRB 211211A and GRB 230307A.

\begin{figure}
	\centering
	\includegraphics[angle=0,scale=0.43]{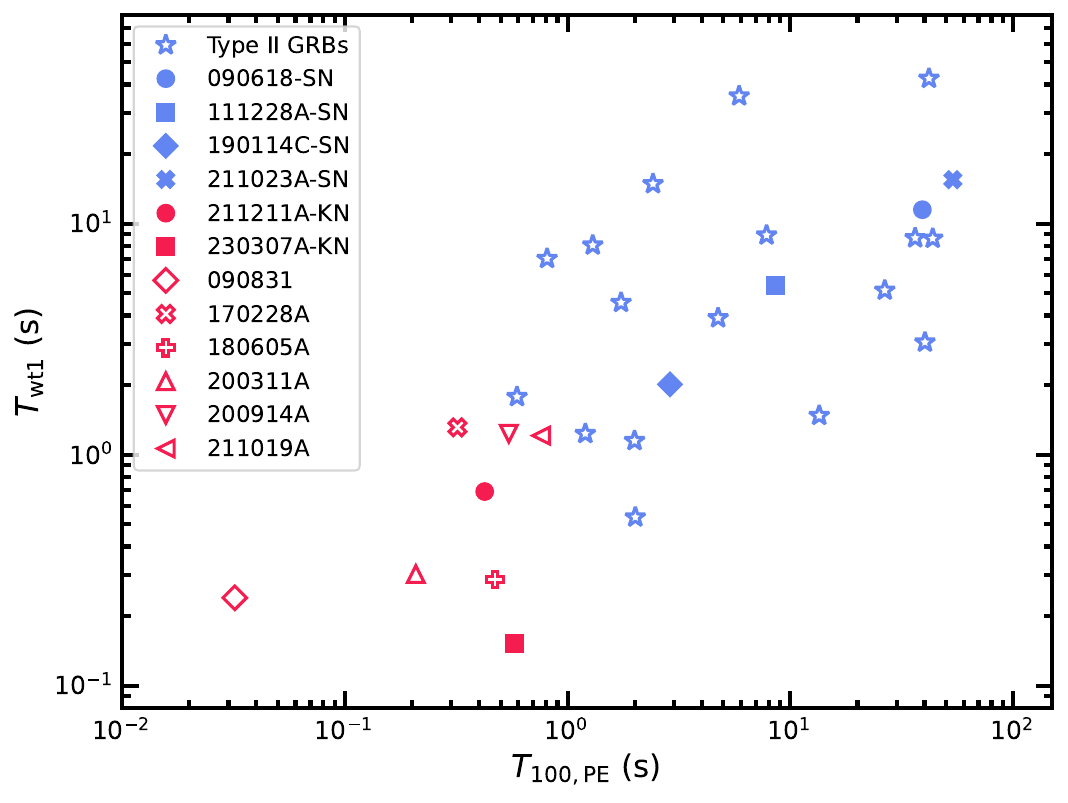}
	\includegraphics[angle=0,scale=0.43]{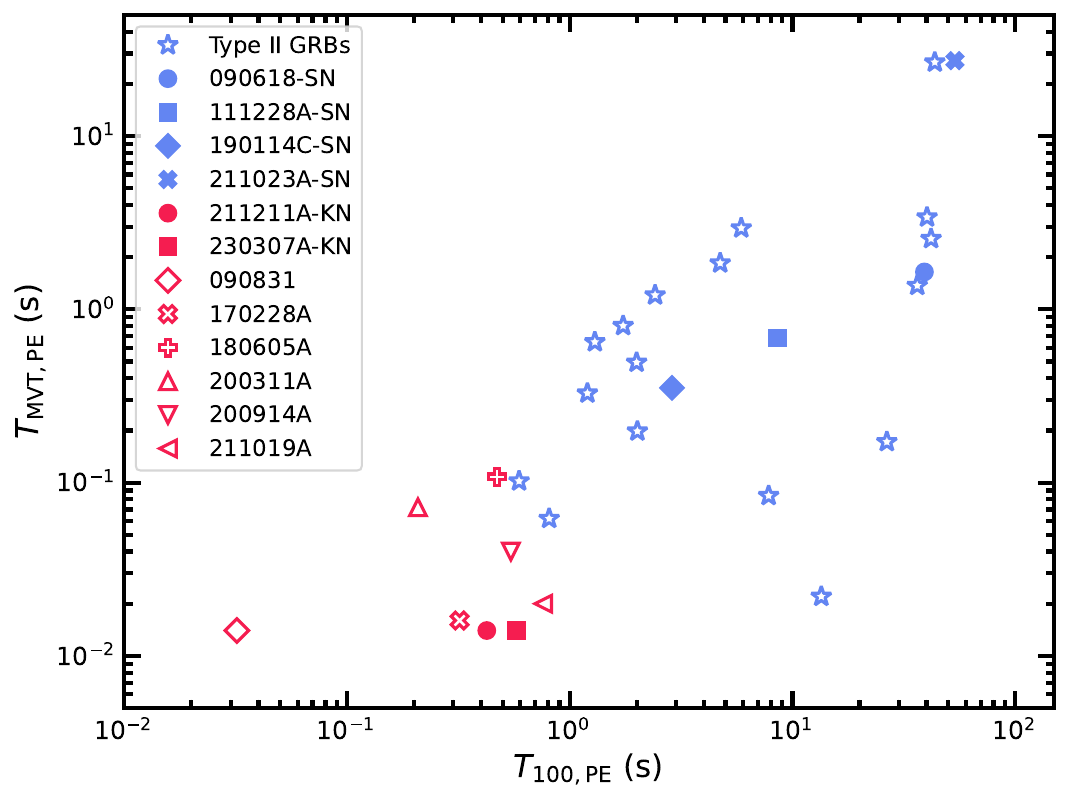}
	\caption{The $T_{\rm 100,PE}$--$T_{\rm wt1}$ and $T_{\rm 100,PE}$--$T_{\rm MVT,PE}$ planes.}
	\label{figure:duration}
\end{figure}

Reviewing the properties of Type IL GRBs, we find that Type IL GRBs appear to be characterized by short $T_{\rm 100,PE}$, $T_{\rm wt1}$, and $T_{\rm MVT,PE}$ and their main emissions also follow the $E_{\rm p,z}$--$E_{\rm iso}$ correlation of Type I GRBs.
The kernel density plots in Figure \ref{figure:par} also support these characteristics.
In addition, $T_{\rm 100,PE}$ and $T_{\rm wt1}$ of Type IL GRBs are consistent with the analysis results of precursor emission in SGRBs \citep{2019ApJ...884...25Z,2020ApJ...902L..42W}.
Although Type IL GRBs and Type II GRBs are not clear boundary in any parameter plane, Type IL GRBs are indeed located in the lower-left region of both the $T_{\rm 100,PE}$--$T_{\rm wt1}$ and $T_{\rm 100,PE}$--$T_{\rm MVT,PE}$ planes and are not contaminated by Type II GRBs, as shown in Figure \ref{figure:duration}.
In addition, we also verify that neither t-SNE nor UMAP can reproduce the above classification results without considering $T_{\rm 100,PE}$, $T_{\rm wt1}$, and $T_{\rm MVT,PE}$, which supports their importance in distinguishing Type IL GRBs from Type II GRBs.
However, we note that they cannot be simply distinguished by a single parameter and that the $E_{\rm p,z}$--$E_{\rm iso}$correlation is also a key attribute that distinguishes them.
For example, GRB 241029A has a short $T_{\rm wt1}$, but its $T_{\rm 100,PE}$ and $T_{\rm MVT,PE}$ are relatively large. 
Furthermore, its main emission does not fall within the Type I GRB populations in the $E_{\rm p,z}$--$E_{\rm iso}$ plane, suggesting a possible collapsar origin.

Although t-SNE and UMAP are powerful nonlinear dimensionality reduction algorithms for visualizing high-dimensional data, they cannot perfectly preserve global distances or data density, and are sensitive to random initialization and hyperparameter choices.
To assess the stability of the machine learning results, we varied key hyperparameters (e.g., setting the $perplexity$ and $n\_neighbors$ to 5, 6, and 7, respectively) and employed different random initializations.
While the embeddings are indeed sensitive to random initialization and the classification structure does not consistently appear under all configurations, we find that in many cases, the separation between the Type IL GRB and Type II GRB clusters remain distinguishable and reproducible within the appropriate parameter range. 
We believe this limitation is largely due to the relatively small sample size in our current dataset and can be verified again when more samples are observed in the future.
Regarding the GMM clustering, we emphasize that it was not applied as a strict clustering framework in this work, but rather as a convenient tool to provide a visual and quantitative interpretation of the visible separation in the t-SNE and UMAP space.
This approach may introduce biases when the embedding space does not fully reflect the original distribution of the data.
Therefore, we also applied other clustering methods, hierarchical density-based spatial clustering of applications with noise \citep[HDBSCAN;][]{2017JOSS....2..205M} and spectral clustering \citep[see][and the references therein]{2023ApJ...945...67S}, which has been successfully applied in previous studies to cluster t-SNE and UMAP embeddings \citep{2023ApJ...945...67S,2023MNRAS.519.1823Z}.
We find that the classification from HDBSCAN and spectral clustering were consistent with those obtained via GMM on both the t-SNE and UMAP embeddings, which enhanced the reliability of the results.
Moreover, the temporal and spectral characteristics of the two clusters show noticeable differences.
In particular, Type IL GRBs share similarities with Type I GRBs, and appear to be distinct from Type II GRBs.
The consistency of results across multiple methods suggests that the Type IL GRB and Type II GRB clusters are not an artifact of specific method and hyperparameters, but rather reflects an intrinsic structure in the data.
We also find that there might be some substructures in the t-SNE and UMAP maps. 
However, some of these substructures may result from the limited sample, so we do not focus too much on these substructures.
Furthermore, the preprocessing step involves extracting a set of 12 temporal and spectral parameters from the GRB light curves. 
While this reduces data complexity and avoids the need for using full light curve time series, it may also omit subtle temporal or spectral structures not captured by the selected parameters.
Future work can consider classification directly based on the light curve to test the universality of our classification.
	
\begin{figure*}
	\centering
	\includegraphics[angle=0,scale=0.37]{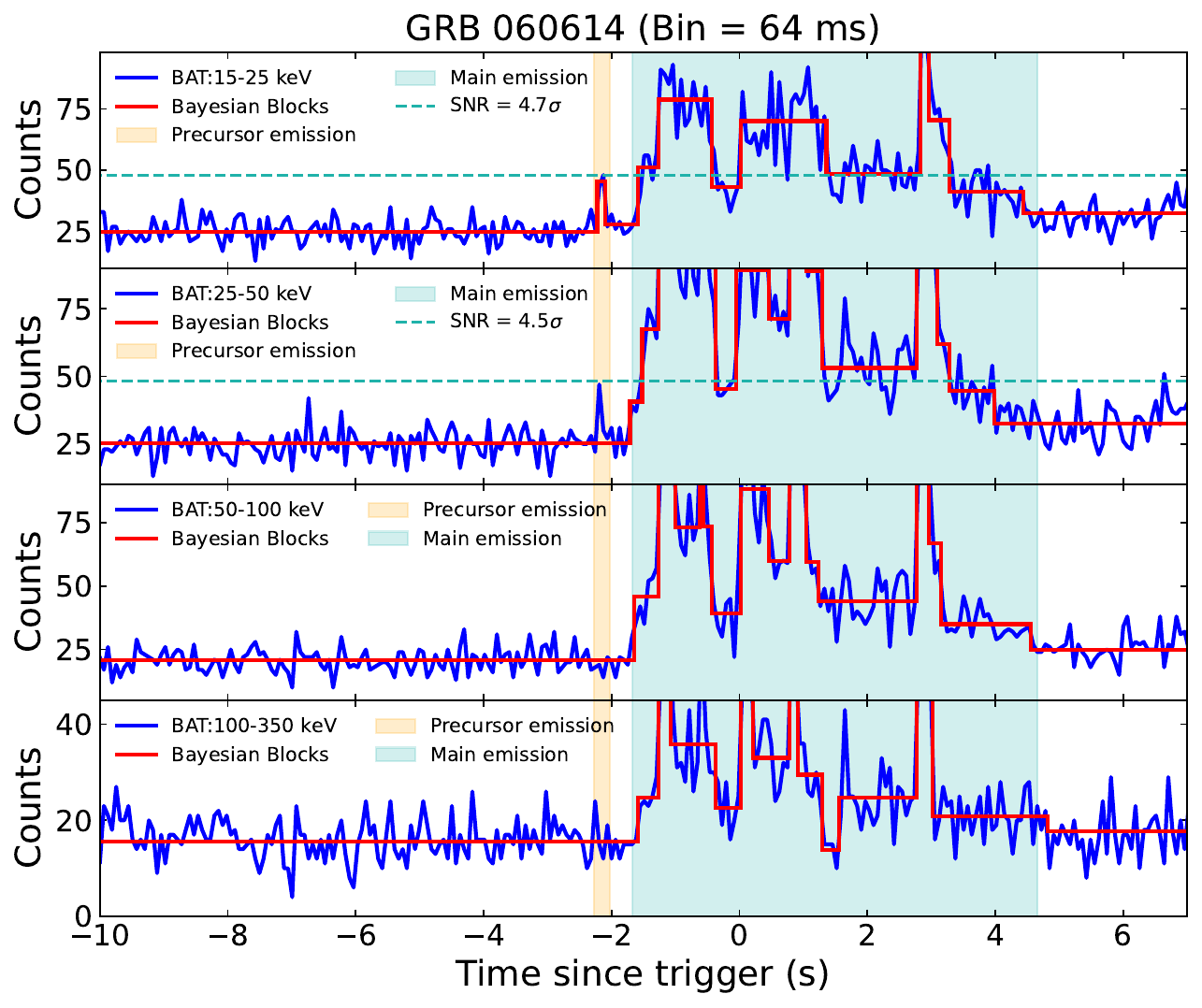}
	\includegraphics[angle=0,scale=0.37]{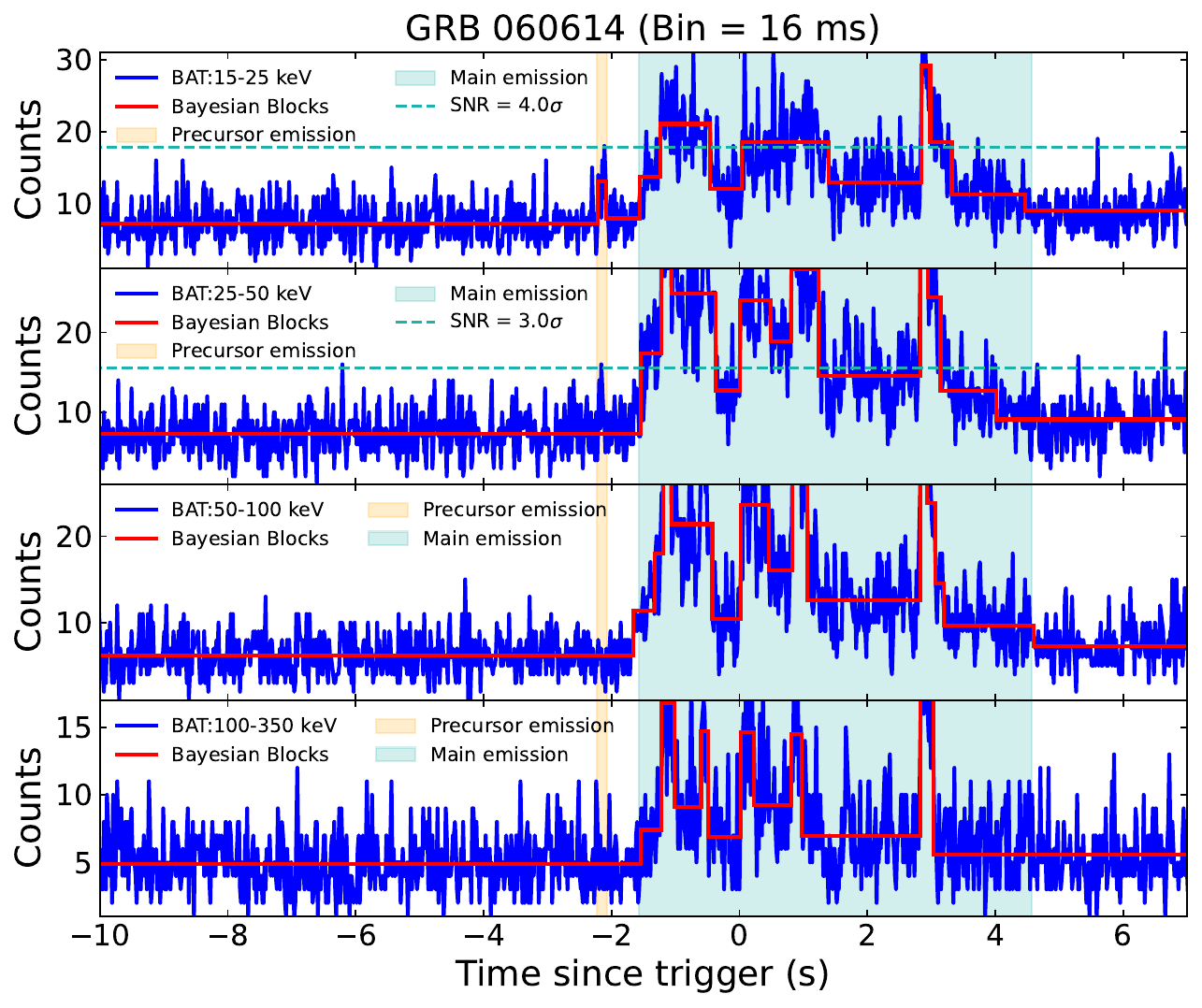}\\
	\includegraphics[angle=0,scale=0.49]{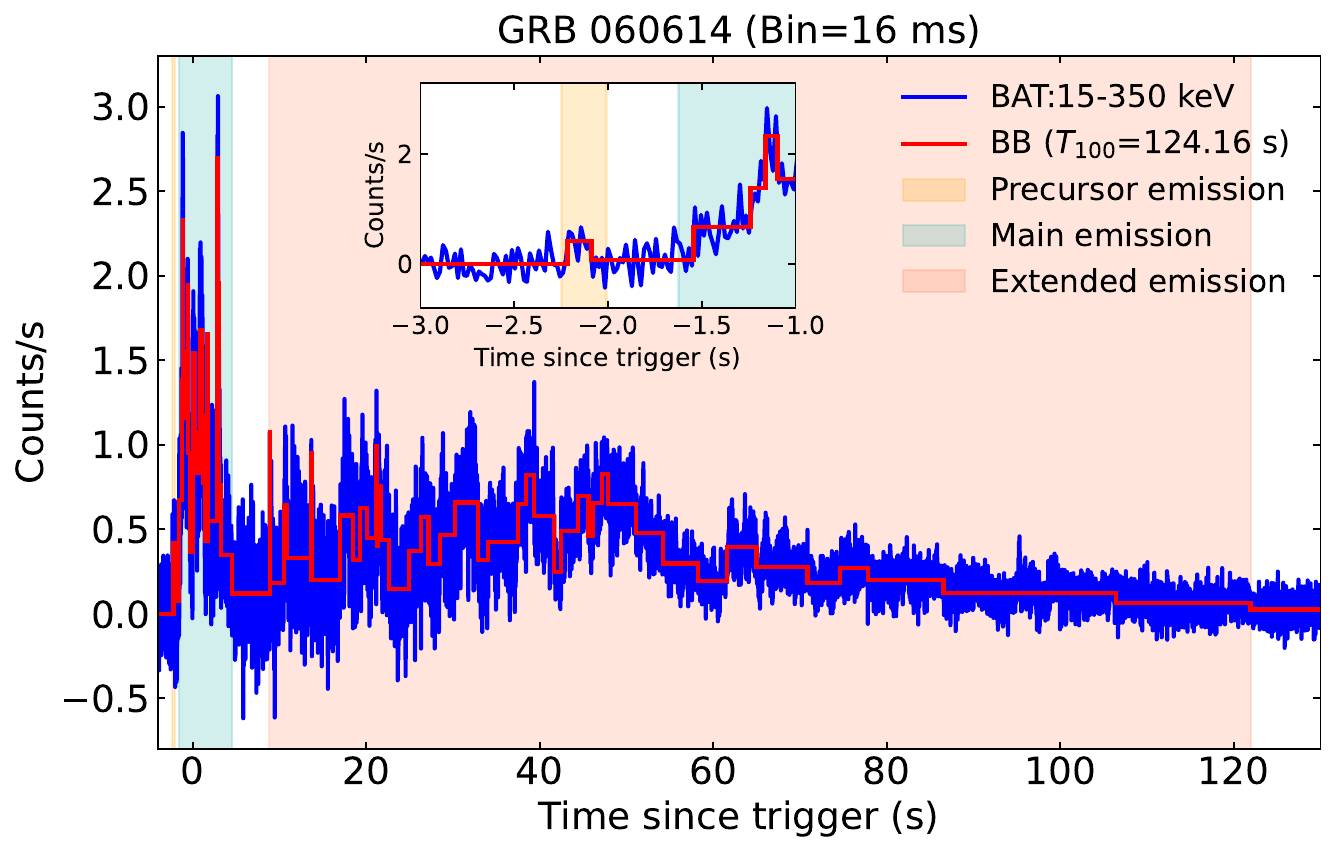}
	\caption{The two figures on the top panel represent light curves of GRB 060614 in the 15--25, 25--50, 50--100, and 100--350 keV bands with 64 ms and 16 ms resolutions, respectively. The green dashed line represents SNR. The bottom plane is the Swift/BAT mask-weighted light curve of GRB 060614 in the 15--350 keV band. }
	\label{figure:060614}
\end{figure*}

GRB 060614 is a peculiar LGRB observed by Swift and Konus, consisting of a $\sim6$ s main emission and a $\sim100$ s extended emission. 
It is the first LGRB believed to originate from a merger, as evidenced by the presence of a KN in its afterglow \citep{2015NatCo...6.7323Y}.
Although it was not detected by Fermi/GBM, it is necessary to examine its consistency with the Type IL GRBs. 
We analyze the BAT data for GRB 060614, and use the standard BAT tool, \texttt{batbinevt}, to extract the light curve for various time resolutions and energy bands.
The light curves of GRB 060614 are shown in Figure \ref{figure:060614}.

Firstly, we identify the precursor emission of GRB 060614 in the 16 ms and 64 ms time-resolution light curves using the Bayesian block method.
The precursor emission is mainly detected in the 15--25 and 25--50 keV bands, with a significance exceeding 4.5 $\sigma$ in the 64 ms time-resolution light curve, while no significant signal was observed in the higher energy band, which is consistent with the typical soft spectrum of precursor emission.
Finally, we confirm the precursor, main, and extended emissions of GRB 060614 in the 15--350 keV band with a time resolution of 16 ms using the Bayesian block method, indicating that it is a three-espiode GRB, and we calculated their $T_{\rm MVT}$ and $\tau$.

GRB 060614 consists of a 0.13 s precursor emission, a 6.08 s main emission, and a 113.07 s extended emission, resulting in a total duration of 124.16 s, with $T_{\rm wt1} = 0.54$ s. 
The spectral lag of the main emission is $\tau_{\rm 42,ME} = 40.93 \pm 7.31$ ms.
For the whole emission, the spectral lag is $\tau_{\rm 42,WE} = 33.39 \pm 5.11$ ms, both comparable to those of the main emission and consistent with Type IL GRBs. 
The MVT of the whole emission occurs in the main emission, only 8 ms, and the MVT of main emission is 72 ms, both consistent with Type IL GRBs.
Its main emission also lies within the range of Type I GRBs in the $E_{\rm p,z}$--$E_{\rm iso}$ plane \citep{2023ApJ...950...30Z}.

To further confirm whether GRB 060614 belongs to Type IL GRBs, we applied t-SNE and UMAP to the sample including GRB 060614, as shown in Figure \ref{figure:classification}. 
The $E_{\rm p}$ values of the main emission and whole emission are taken from \cite{2023ApJ...950...30Z}, $E_{\rm p,ME} = 302$ keV and $E_{\rm p,WE} = 76$ keV, respectively. 
Obviously, GRB 060614 together with the 8 GRBs mentioned above forms the Type IL GRB population.
These results further support our conclusion that Type IL GRBs originate from mergers and can be distinguished from typical Type II GRBs using machine learning methods.

Note that instrumental selection effects are an important consideration when extending our method beyond the Fermi/GBM dataset. Although GRB 060614 was not observed by Fermi, it represents a particularly suitable case study due its confirmed merger origin; this also inevitably introduces potential instrumental biases. Nevertheless, we emphasize that our analysis of GRB 060614 is primarily revelatory, aimed at verifying whether this Type IL GRB exhibits a precursor emission (a feature not previously confirmed), and whether its precursor properties are consistent with those of other Type IL GRBs observed by Fermi. The BAT data for GRB 060614 clearly reveal a precursor emission, which is consistent with the characteristics observed in Type IL GRBs detected by Fermi. This consistency suggests that our classification may be relatively insensitive to instrumental selection effects, possibly due to the typically soft nature of precursor emissions, which fall within the overlapping low-energy sensitivity ranges of Fermi/GBM and Swift/BAT. Meanwhile, we point out that it is necessary to consider the instrument selection effect in similiar, future classification works.

\section{discussion} \label{sec:discussions}

Due to the relative weakness of the precursor emission and its low detection rate, our understanding of it remains insufficient.
Recently, \cite{2025ApJ...979...73W} proposed that short $T_{\rm wt1}$ may be the most significant feature distinguishing Type IL GRBs from Type II GRBs, as shorter quiescent episodes and longer durations cause Type IL GRBs to form a distinct cluster on the $T_{\rm wt1}$--$T_{90}$ plane.
However, we do not observe such a clear boundary, which may be due to a selection effect, as their sample do not include GRBs with $T_{\rm wt1}$ values bridging the two populations.
Nevertheless, we also find that Type IL GRBs are indeed located in the lower-left region of the $T_{\rm wt1}$--$T_{\rm 100,PE}$ plane and are not contaminated by Type II GRBs.
The same holds true for the $T_{\rm 100,PE}$--$T_{\rm MVT,PE}$ planes. 
Therefore, we suggest that precursor emission plays a crucial role in distinguishing Type IL and Type II GRBs, with the short $T_{\rm 100,PE}$, $T_{\rm wt1}$, and $T_{\rm MVT,PE}$ serving as key characteristics of Type IL GRBs.

The precursor emission is generally soft and short, occurring prior to the main emission \citep{1995ApJ...452..145K,2005MNRAS.357..722L}.
Although some studies suggest that precursor emission and main emission share a common origin \citep{2008ApJ...685L..19B,2009A&A...505..569B,2010ApJ...723.1711T,2014ApJ...789..145H,2015MNRAS.448.2624C,2021ApJS..252...16L,2022ApJ...928..152L,2024ApJ...970...67D}, others suggest that they are independent of each other \citep{1995ApJ...452..145K,2005MNRAS.357..722L,2018NatAs...2...69Z,2019ApJS..242...16L,2019ApJ...884...25Z}.
Theoretically, several models have been proposed to explain the physical origins of the precursor emission.
For Type IL GRBs, the precursor emission can occur prior to the merger. 
There are two main branches in the pre-merger models: the crustal breaking of NS, includes the resonant shattering flare \citep[RSF;][] {2012PhRvL.108a1102T,2020PhRvD.101h3002S,2022MNRAS.514.5385N} and magnetar super flare \citep[MSF;][] {2022ApJ...939L..25Z} models, and the interaction between magnetospheres \citep{2013PhRvL.111f1105P,2018ApJ...868...19W}.
Given that the precursor emissions of Type IL GRBs with redshift have extremely high energy and luminosity, with $L_{\rm iso,PE} \sim 10^{49}$--$10^{50}$ erg s$^{-1}$, a key issue for discussion is whether these models can provide sufficient energy.

In general, the maximum luminosity of RSF and MSF models is about $10^{47}$ erg s$^{-1}$, which is extracted from the crust or core through strong magnetic fields, as estimated by
\begin{equation}\label{Lmax}
	{L}_{\max }\sim {10}^{47}\ \mathrm{erg}\cdot {{\rm{s}}}^{-1}\left(\displaystyle \frac{v}{c}\right){\left(\displaystyle \frac{{B}_{\mathrm{surf}}}{1{0}^{13}\ {\rm{G}}}\right)}^{2}{\left(\displaystyle \frac{{R}}{10\ \mathrm{km}}\right)}^{2}\\,
\end{equation}
where $v$ is the velocity of the perturbations, $R$ is the radius of NS , and $B_{\rm surf}$ is the local surface field strength \citep{2013ApJ...777..103T}.
Only when the surface magnetic field of NS reaches $10^{14}$--$10^{15}$ G can sufficient energy be provided, which is the typical magnetic field of a magnetar, indicating that the binary system contains a magnetar \citep{2017ARA&A..55..261K}.
Additionally, under a strong magnetic field ($> 10^{15}$ G), the luminosity produced by the magnetosphere interaction model can also reach $10^{50}$--$10^{51}$ erg s$^{-1}$, although it is slightly lower than that of the RSF and MSF models \citep{2012ApJ...757L...3L,2013PhRvL.111f1105P}.

All these pre-merger models require at least one neutron star in the binary system to have an extremely strong magnetic field, $B_{\rm surf} \sim 10^{15}$ G, in order to achieve $L_{\rm iso,PE} \sim 10^{50}$ erg s$^{-1}$, which indicates the presence of a magnetar with a strong magnetic field in the binary system \citep{2023ApJ...954L..29D}.

The unique observation property of the binary compact merger is the time delay between the GRBs and the GW generated at the end of the merger ($\Delta t$).
Therefore, the pre-merger models must also consider whether their timescales are compatible with GW observations. 
Both the RSF and MSF models predict that precursor emission can occur within 0.1 s prior to the merger. 
Thus, if the precursor emission indeed occurs before the merger and is not delayed by other physical processes, the time delay between the gravitational wave and the main emission should be comparable to $T_{\rm wt1}$.
The Fermi/GBM observations of GRB 170817A, which originated from the binary neutron star merger and associated with GW170817, show a $\Delta t \sim 1.7$ s delay \citep{2017ApJ...851L..18W}. 
Similarly, the potential GRB 150914A associated with GW150914 has a $\Delta t \sim 0.4$ s \citep{2016ApJ...826L...6C}, and subthreshold GRB, GBM-190816, associated with a subthreshold LIGO/Virgo compact binary merger candidate has a $\Delta t \sim 1.57$ s \citep{2020ApJ...899...60Y}.
\cite{2016ApJ...827...75L} predicted that for NS-NS and NS-BH mergers, $\Delta t$ ranges from 0.01 s to a few seconds. 
Although the distribution of $\Delta t$ is still uncertain, the short $T_{\rm wt1}$ for Type IL GRBs falls within the $\Delta t$ range.

Although the physical model of the precursor emission remains uncertain, if it indeed plays a key role in distinguishing Type IL and Type II GRBs, pre-merger models are our preferred choice, as they can provide sufficient energy and are consistent with the timescale of GW observations.
Previously, \cite{2020PhRvD.102j3014C} found that $T_{\rm wt1}$ exhibits a bimodal distribution with peaks of 0.54 s and 32 s, respectively, which also indicates that some precursor emissions might have been produced before the merger.

\section{conclusions} \label{sec:conclusions}

Recent observations of GRB 211211A and GRB 230307A suggest the possible existence of a subclass of LGRBs originated from mergers, Type IL GRBs, challenging the traditional understanding of GRBs.
However, apart from detecting KNe in their optical afterglows, no rapid identification method has been established, and observational limitations hinder its applicability to most GRBs.
In this paper, we propose a method based solely on prompt emission to rapidly identify Type IL GRBs and provide new insights into their nature.

A common feature of GRB 211211A and GRB 230307A is their three-episode light curve structure: precursor emission, main emission, and extended emission.
We comprehensively searched the Fermi Catalog and identified 29 three-episode GRBs from 3883 GRBs.
We find that both mergers and collapsars can produce such three-episode GRBs.
Based on 12 parameters, we apply t-SNE and UMAP to the three-episode GRBs, and both the t-SNE and UMAP maps clearly display two clusters. 
Interestingly, we find that the two GRBs associated with KN and the four GRBs associated with SN belong to different clusters, suggesting that these clusters may correspond to GRBs with different origins, Type IL GRBs and Type II GRBs.
Type IL GRBs consist of eight GRBs: GRB 090831, GRB 170228A, GRB 180605A, GRB 200311A, GRB 200914A, GRB 211019A, GRB 211211A, and GRB 230307A.

We carefully examined temporal and spectral properties of Type IL GRBs and found that they are indeed very similar to Type I GRBs, such as relatively small $\tau$ and MVT, as well as their main emission deviating from the Type II GRB populations in the $E_{\rm p,z}$--$E_{\rm iso}$ plane.
These results suggest that machine learning may be capable of distinguishing LGRBs with different origins and that there exists a subclass of Type I GRBs, i.e. Type IL GRBs.
They exhibit temporal and spectral properties similar to Type I GRBs, yet their whole emission duration is significantly longer than that of typical Type I GRBs. 
Furthermore, $T_{\rm 100,ME}$ can be very short or extend beyond 10 s, posing new challenges to GRB theoretical models.

We find that Type IL GRBs are characterized by short $T_{\rm 100,PE}$, $T_{\rm wt1}$, and $T_{\rm MVT,PE}$.
Therefore, we suggest that precursor emission plays a crucial role in distinguishing Type IL and Type II GRBs.
To further verify our hypothesis, we analyze the confirmed Type IL GRB, GRB 060614, and, for the first time, identify its high-significant precursor emission with short $T_{\rm 100,PE}$, $T_{\rm wt1}$, and $T_{\rm MVT,PE}$. 
The machine learning results suggest that it is also classified as Type IL GRB.
We conclude the characteristics of Type IL GRBs as follow:
\begin{enumerate}
	\item Consists of precursor, main, and extended emissions. It cannot be ruled out that the precursor and extended emissions of some GRBs were not observed because of their weakness.
	
	\item The short $T_{\rm 100,PE}$, $T_{\rm wt1}$, and $T_{\rm MVT,PE}$.
	
	\item The main emission follows the $E_{\rm p,z}$--$E_{\rm iso}$ correlation of the Type I GRBs. 
\end{enumerate}

Except for GRB 211211A and GRB 230307A, the remaining six Type IL GRBs have not received attention and have been buried in the vast amount of data. 
In fact, more special GRBs may already have been observed, but have not drawn attention. 
This can be attributed to two factors. 
Firstly, the limitations of instruments may prevent rapid localization, making multi-wavelength follow-up observations challenging. Fortunately, the launch of the Space-based multi-band astronomical Variable Objects Monitor (SVOM) satellite and Einstein Probe (EP) will allow for faster and more precise detection of GRBs, enabling further confirmation of special GRB samples through follow-up multi-wavelength observations, which will advance the study of GRBs. 
Secondly, it is difficult to rapidly determine whether a GRB is special and worth further tracking. 
This paper proposes a rapid identification method that may facilitate the discovery and tracking of more such special GRBs.

\section*{Acknowledgements}
We thank the anonymous reviewers for their insightful comments/suggestions.
We acknowledge the use of public data and software provided by the Fermi Science Support Center and the UK Swift Science Data Centre at the University of Leicester. 
This work was supported in part by the National Natural Science Foundation of China (No. 12273122), and National Astronomical Data Center, the Greater Bay Area, under 2024B1212080003.
F.-W.Z. acknowledges the support from the National Natural Science Foundation of China (No. 12463008). 

\section*{Data Availability}
The data underlying this article are available in the article and in its online supplementary material.
The Fermi GBM data underlying this article are publicly available at the FTP website (https://heasarc.gsfc.nasa.gov/FTP/fermi/data/gbm).
The Swift BAT data underlying this article are publicly available at https://www.swift.ac.uk.



\bibliographystyle{mnras}
\bibliography{ref} 


\bsp	
\label{lastpage}

\end{document}